\documentclass[aps,pre,color,psfig,epsf,notitlepage]{revtex4-1}
\usepackage{graphicx} 
\usepackage{color}
\usepackage{amsmath}
\usepackage{amsfonts}
\usepackage{amssymb}
\usepackage{enumitem}
\usepackage{epsf}
\usepackage{epstopdf}
\usepackage{enumitem}
\usepackage{babel}
\usepackage{txfonts}

\usepackage{tikz}
\usepackage{tikz-3dplot}
\usetikzlibrary{automata}
\usetikzlibrary{arrows}
\usetikzlibrary{positioning,calc}
\usetikzlibrary{graphs}
\usetikzlibrary{graphs.standard}
\usetikzlibrary{arrows,decorations.markings}
\usepackage{tkz-graph}
\usetikzlibrary{chains,fit,shapes}
\usetikzlibrary{calc}

\tikzset{every loop/.style={min distance=10mm,looseness=10}}
\tikzset{every state/.style={minimum size=2mm}}

\usepackage[T1]{fontenc}
\usepackage[babel=true]{microtype}
\usepackage{amsmath}
\usepackage{amsfonts}
\usepackage{amssymb}
\usepackage{bm}
\usepackage{esint}%

\let\originalleft\left
\let\originalright\right
\renewcommand{\left}{\mathopen{}\mathclose\bgroup\originalleft}
\renewcommand{\right}{\aftergroup\egroup\originalright}
\def\bignicefrac#1#2{
    \left. {%
    \raise1.5ex\hbox{$\displaystyle#1$}%
    }%
    \kern-.5em%
    \middle/%
    \kern-.35em%
    {%
    \lower1.25ex\hbox{$\displaystyle#2$}%
    } \right.%
    }

\newcommand{\deriv}[2]{\frac{\mathrm{d}#1}{\mathrm{d}#2}}

\newcommand{\hderiv}[3]{\frac{\mathrm{d}^#1 #2}{\mathrm{d}#3^#1}}%              higher derivatives

\newcommand{\nth}{\textsuperscript{th}~}

\newcommand\boltz{{k_\text{B}}}%                                                Boltzmann constant
\newcommand\bert{{\mathscr{B}}}%                                                broken symmetry variable
\newcommand\epair{{E_\text{pair}}}%                                             pairing energy LAMMPS output
\newcommand\epairNorm{{e_\text{pair}}}%                                         normalized pairing energy from LAMMPS
\newcommand{\av}[1]{\left\langle#1\right\rangle}

\makeatletter
\newcommand{\size}[1]{\bBigg@{#1}}
\makeatother

\newcommand{\ve}[1]{\bm{#1}}%                                                   ISO-conformal
\usepackage{isomath}%                                                           switch off to change conventions
\usepackage{graphicx}
\usepackage{tikz}
\usepackage{tikz-3dplot}
\usepackage{pgfplots}
\pgfplotsset{compat=newest}%                                                    \pgfplotsset{compat=1.10}

\pgfplotsset{
    CI1/.style={
        legend image code/.code={%
            \node[anchor=center,fill=light-gray, opacity=0.25] at (0.3cm,0cm) {};
        }
    },
    CI2/.style={
        legend image code/.code={%
            \node[anchor=center,fill=yellow, opacity=0.5] at (0.3cm,0cm) {};
        }
    },
    CI3/.style={
        legend image code/.code={%
            \node[anchor=center,fill=blue, opacity=0.25] at (0.3cm,0cm) {};
        }
    },
    CI4/.style={
        legend image code/.code={%
            \node[anchor=center,fill=red, opacity=0.25] at (0.3cm,0cm) {};
        }
    },
}
\usepackage{readarray}
\usepackage{tkz-graph}
\GraphInit[vstyle = Simple]
\tikzset{
    VertexStyle/.style={shape=coordinate}%                                      tex.stackexchange.com/questions/265780
    }
\newcommand{\loopNarrowN}[1]{
    \Loop[
        dist = 2cm,
        dir = NO,
        style={
            out=67.5,
            in=112.5
            }
        ](#1)%
}
\newcommand{\loopNarrowS}[1]{
    \Loop[
        dist = 2cm,
        dir = SO,
        style={
            out=-67.5,
            in=-112.5
            }
        ](#1)%
}
\newcommand{\loopNarrowNE}[1]{
    \Loop[
        dist = 2cm,
        dir = SOWE,
        style={
            out=22.5,
            in=67.5
            }
        ](#1)%
}
\newcommand{\loopNarrowNW}[1]{
    \Loop[
        dist = 2cm,
        dir = SOWE,
        style={
            out=112.5,
            in=157.5
            }
        ](#1)%
}
\newcommand{\loopNarrowSW}[1]{
    \Loop[
        dist = 2cm,
        dir = SOWE,
        style={
            out=202.5,
            in=247.5
            }
        ](#1)%
}
\newcommand{\loopNarrowSE}[1]{
    \Loop[
        dist = 2cm,
        dir = SOWE,
        style={
            out=292.5,
            in=337.5
            }
        ](#1)%
}
\newcommand{\loopNarrowWSW}[1]{
    \Loop[
        dist = 2cm,
        dir = SOWE,
        style={
            out=225,
            in=180
            }
        ](#1)%
}
\newcommand{\loopNarrowENE}[1]{
    \Loop[
        dist = 2cm,
        dir = NOEA,
        style={
            out=0,
            in=45
            }
        ](#1)%
}
\newcommand{\loopNarrowESE}[1]{
    \Loop[
        dist = 2cm,
        dir = SOEA,
        style={
            out=0,
            in=-45
            }
        ](#1)%
}
\newcommand{\loopNarrowE}[1]{
    \Loop[
        dist = 2cm,
        dir = EA,
        style={
            out=-22.5,
            in=22.5
            }
        ](#1)%
}
\newcommand{\loopNarrowW}[1]{
    \Loop[
        dist = 2cm,
        dir = EA,
        style={
            out=157.5,
            in=202.5
            }
        ](#1)%
}

\newcommand{\loopLongS}[1]{
    \Loop[
        dist = 3cm,
        dir = SO,
        style={
            out=-67.5,
            in=-112.5
            }
        ](#1)%
}
\newcommand{\loopLongNE}[1]{
    \Loop[
        dist = 3cm,
        dir = SOWE,
        style={
            out=22.5,
            in=67.5
            }
        ](#1)%
}

\newcommand{\s}{\hspace{1em}}%                                                  spacer command to align diagrams in table TAG: JKL
\usepackage{grffile}
\usepackage{siunitx}
\DeclareSIUnit \kT { \text { \ensuremath { \boltz T } } }
\usepackage{nicefrac}

\usepackage{titletoc}

\definecolor{light-gray}{gray}{0.75}
\definecolor{light-light-gray}{gray}{0.875}
\colorlet{dark-green}{green!70!black}
\usetikzlibrary{arrows.meta,bending,decorations.markings,intersections,automata,pgfplots.units,spy,circuits,circuits.ee.IEC,calc,math,patterns}
\usepgfplotslibrary{fillbetween,decorations.softclip,external}
\begin{document}

\title{Combinatorics and topological weights of chromatin loop networks}

\author{Andrea Bonato$^{1}$, Dom Corbett$^2$, Sergey Kitaev$^3$, Davide Marenduzzo$^2$, Alexander Morozov$^2$, Enzo Orlandini$^4$}
\affiliation{$^1$ Department of Physics, University of Strathclyde, Glasgow G4 0NG, Scotland, United Kingdom \\
$^2$SUPA, School of Physics and Astronomy, The University of Edinburgh, Edinburgh, EH9 3FD, Scotland, United Kingdom \\
$^3$ Department of Mathematics and Statistics, University of Strathclyde, Glasgow, G1 1XH, United Kingdom \\
$^4$Department of Physics and Astronomy, University of Padova and  INFN, Sezione Padova,
Via Marzolo 8, I-35131 Padova, Italy}

\begin{abstract}
Polymer physics models suggest that chromatin spontaneously folds into loop networks with transcription units (TUs), such as enhancers and promoters, as anchors. Here we use combinatoric arguments to enumerate the emergent chromatin loop networks, both in the case where TUs are labelled and where they are unlabelled. We then combine these mathematical results with those of computer simulations aimed at finding the inter-TU energy required to form a target loop network. We show that different topologies are vastly different in terms of both their combinatorial weight and energy of formation. We explain the latter result qualitatively by computing the topological weight of a given network -- i.e., its partition function in statistical mechanics language -- in the approximation where excluded volume interactions are neglected. Our results show that networks featuring local loops are statistically more likely with respect to networks including more non-local contacts. We suggest our classification of loop networks, together with our estimate of the combinatorial and topological weight of each network, will be relevant to catalogue 3D structures of chromatin fibres around eukaryotic genes, and to estimate their relative frequency in both simulations and experiments. 
\end{abstract}

\maketitle

\section{Introduction}

%simple model for active chromatin as loop network
Chromatin is a protein-DNA composite polymer that provides the building block of chromosomes, and it constitutes the form in which genomic information is stored in the nuclei of eukaryotic cells. Chromatin also provides the genomic substrate for fundamental intracellular processing of DNA, such as transcription and replication~\cite{Calladine1997,Alberts2014}. Long-standing observations suggest that the 3D structure of chromatin is functionally important: for instance, it is known that the 3D structure of a gene locus correlates with its transcriptional activity~\cite{Chiang2022b}. 

Polymer models to determine chromatin structure in 3D are therefore important in this field, and several coarse-grained potentials have been developed to describe them (see, e.g.,~\cite{Rosa2008,Barbieri2012,Jost2014,DiPierro2016,Chiariello2016}, and~\cite{Brackley2020,Chiang2022} for a review of some of these). Typically, coarse-grained polymer models view chromatin as a copolymer, or heterogeneous polymer, where different beads may have different properties to reflect, among others, the local sequence and post-translational modification in DNA-binding histone proteins, such as acetylation or methylation (see, e.g.,~\cite{Jost2014,Chiang2019,Chiang2022b}).

\begin{figure}
\begin{center}
\includegraphics[scale=0.3]{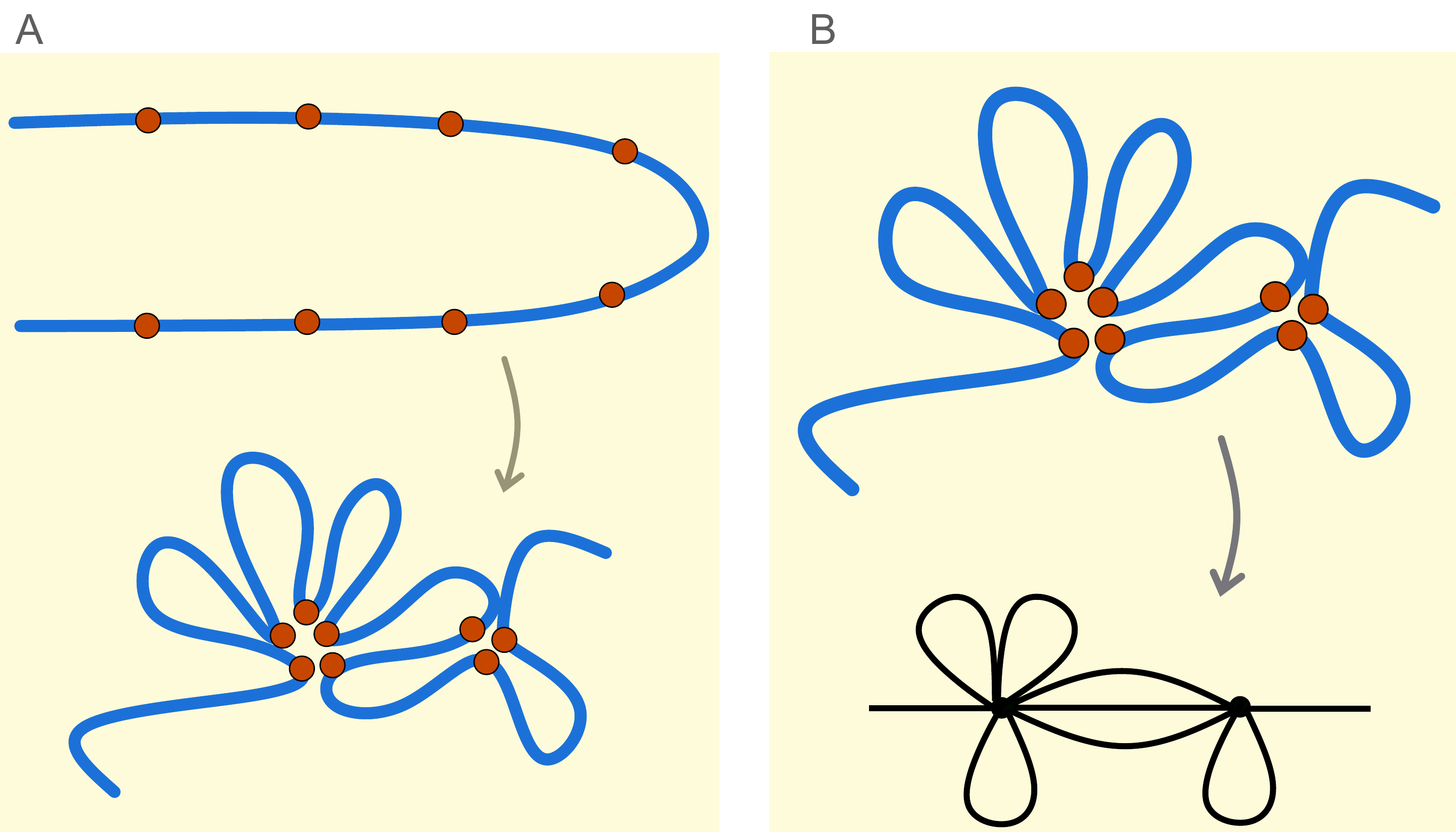}
\caption{(A) (Top) A chromatin fibre with $n=8$ TUs. (Bottom) A possible  structure formed when TUs attract each other, for instance effectively due to the bridging-induced attraction~\cite{Brackley2016}. The structure is made of two clusters and 4 local loops (B) The loop network topology corresponding to this configuration (repeated at the top for clarity) is shown at the bottom of the panel.}\label{fig1}
\end{center}
\end{figure}

A simple copolymer model for active chromatin~\cite{Marenduzzo2009}, which is relevant to our current work, views the fibre as a semiflexible polymer with interspersed ``transcription units'' (TUs, the red circles in Fig.~\ref{fig1}), representing open chromatin regions such as enhancers or promoters which have high affinity for multivalent chromatin-binding proteins associated with  transcription -- such as RNA polymerases and transcription factors, or protein complexes including both of these~\cite{Brackley2016,Brackley2021}. 

%emergence of chromatin loop networks -- JSTAT and BIPS

Simulations of more sophisticated polymer models, resolving chromatin-binding proteins, show that TUs come together due to the bridging-induced attraction, a positive-feedback loop associated with multivalent chromatin-protein binding~\cite{Brackley2016}. The bridging-induced attraction leads to microphase separation into clusters of TUs (and their associated proteins) because clustering the TUs create loops whose entropy grows superlinearly with TU number, eventually balancing the energetic gain of clustering~\cite{Marenduzzo2009,Brackley2016}. This phenomenon provides a mechanistic model for the formation of transcription factories in mammalian nuclei~\cite{Cook2018}. This discussion suggests that in a simpler effective model, one can consider the TUs themselves as sticky for each other, and this is the model sketched in Fig.~\ref{fig1}.

%it is desirable to classify the emerging topologies and find their associated statistical weight

In the copolymer model of Fig.~\ref{fig1}, chromatin loop networks emerge in a steady state due to the sticky nature of TUs. Some natural questions then arise, namely how to classify the emerging network topologies (such as the one in Fig.~\ref{fig1}B), and what the statistical likeliness of observing each of such topologies is. A possible way to classify the loop topologies is by computing the entropic exponent associated with the network, as in~\cite{Duplantier1989}. However, the issue arises that all networks with the same number of nodes and edges (or legs) emanating from each node would have the same entropic exponent, as they have the same number of nodes and edges~\cite{Duplantier1989}. As shown in the companion paper~\cite{short}, simulations suggest, instead, that the probability of observing different networks is not constant at all, so it would be desirable to go beyond the calculation of the entropic exponent and estimate the statistical weight associated with each loop topology.

%definition of labelled and unlabelled chromatin loop networks

We consider two possible classes of chromatin loop networks. First, ``labelled'' networks are those in which the TUs are numbered. This is often relevant in biological examples where different TUs correspond to different regulatory elements, and it may be important in practice to distinguish networks with the same topology and distribution of clusters, but where different TUs participate in the clusters.

\begin{figure}
\begin{center}
\includegraphics[scale=0.3]{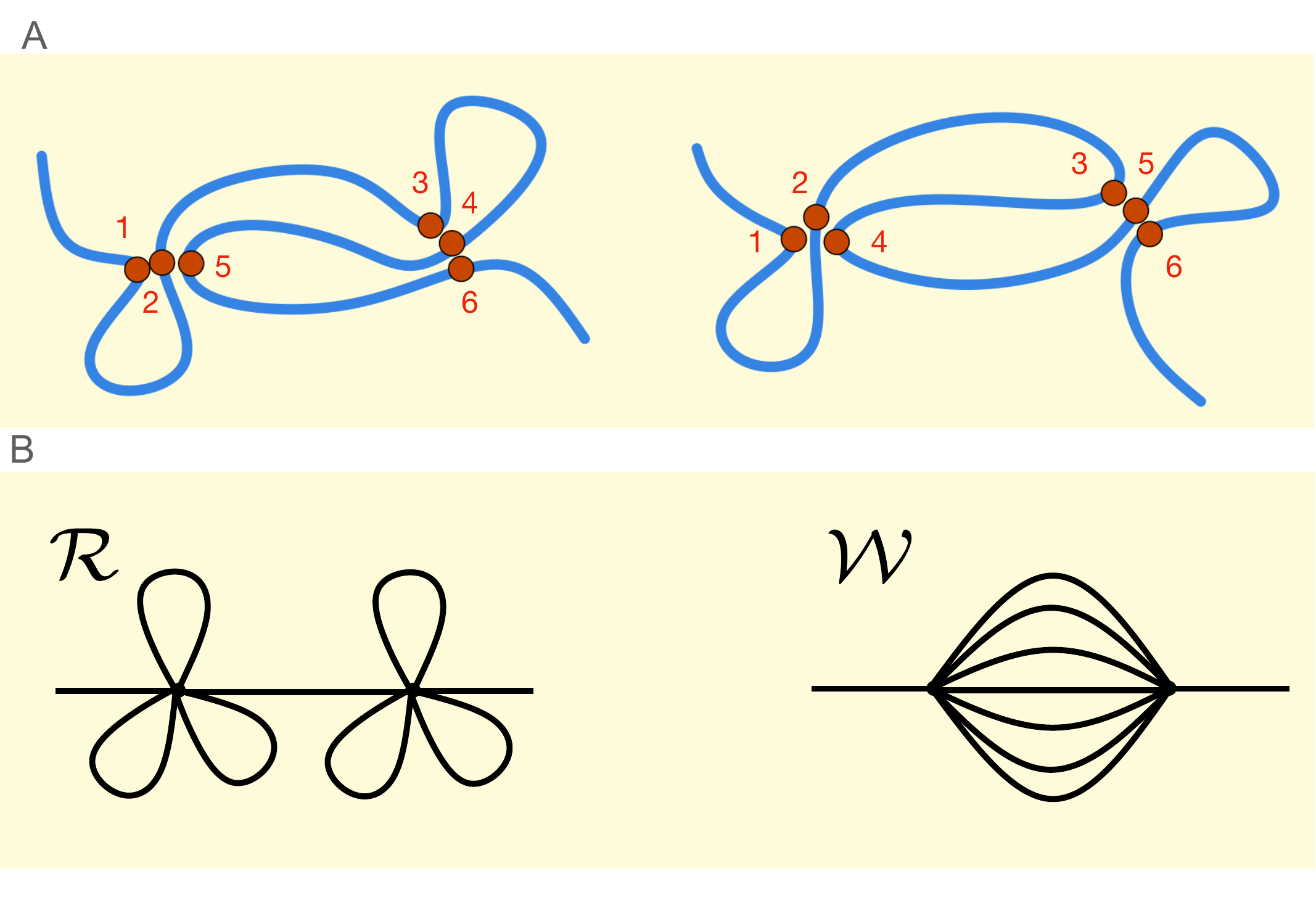}
\caption{(A) An example of two different labelled networks with two clusters which yield the same unlabelled topology (neglecting singletons). (B) Rosette and watermelon topologies. }\label{fig2}
\end{center}
\end{figure}

Second, ``unlabelled'' networks are those where TUs are not numbered, such that different configurations are topologically non-equivalent configurations of our chromatin fibre. For example, the two networks in Fig.~\ref{fig2}A are different labelled networks but represent the same topology when counting unlabelled networks. 
Unlabelled networks are relevant when considering generic topologies, for example, the rosette and watermelon ones in Fig.~\ref{fig2}B, and asking which topology is most often found in gene loci genome-wide. Labelled networks are a lot simpler to count combinatorically with respect to unlabelled ones; this is because it is hard in general to count the multiplicity of labelled networks corresponding to a unique unlabelled network topology.

%structure of the work - first combinatorics (first labelled, then unlabelled), then simulations of copolymer model with different target topologies, then computation of statistical weights of a generic chromatin loop network (first multiplicity/combinatorial weight, then topological weight, statistical weight = product of the two weights)

In the present work, we aim to classify topologies of chromatin loop networks, counting them and finding their statistical weights, which measures the probability of observing them in a polymer model. Our article is structured as follows. First, in Section II we provide combinatorial formulas to count labelled networks. As we shall see, the theory of Bell numbers and partitions provides a powerful way to count such networks. We also find a series of recursion relations which constrain the number of labelled networks with specific properties (e.g., without or with singletons). These recursions are associated with an exponential network-generating function which we find explicit formulas for. Second, in Section III we discuss the case of unlabelled, topologically inequivalent, networks, and derive a formula to count the number of such structures with two clusters, which is of interest in applications to chromatin structures in real gene loci. Section IV contains numerical results obtained by simulating chromatin folding within a specific polymer model, viewing the chromatin fibre as a semiflexible self-avoiding chain with equally spaced sticky sites (the TUs). Here we show that different target topologies require different interaction energies between the TUs to form so that they are in general associated with a different entropic cost of formation. These results complement those discussed in the companion paper~\cite{short}, which show that rosette-like topologies, that are rich in local loops, are much more favoured statistically with respect to others with non-local loops. In Section V, we compute the statistical weight of a generic topology in the simplified case of a phantom freely-jointed chain (i.e., without excluded volume interactions). We show that the weights we compute, although approximate due to the neglect of excluded volume effects, are sufficient to recapitulate the much enhanced statistical likeliness of forming rosette-like networks, in spite of the fact that the combinatoric multiplicities of other topologies are often larger. Finally, Section VI contains our conclusion. 

\section{Combinatorics of labelled chromatin loop networks}

%\section{Labelled chromatin loop networks: combinatorial formulas and generating function}

%While originally, only topologically non-equivalent configurations of the gene locus formed by Transcription Units (TUs), some of which are grouped (or glued) together into {\em clusters} (or {\em sticky sites}), were of interest, it is also useful ({\bf WHY?}) to consider/count configurations under various settings/restrictions where TUs are distinguishable, i.e. {\em labelled}. 

We first consider the case of labelled chromatin loop networks, where more progress can be done analytically. Therefore, in this Section TUs are assumed to be labelled from $1$ to $n$, and we think of TUs as the set $\{1,2,\ldots,n\}$ that is also denoted by $[n]$. %Additionally, unless otherwise specified, we let $a_n$ denote the number of configurations in question, and $A(t)=\sum_{n\geq 0}a_nt^n$ the (ordinary) corresponding  generating function (g.f.).  

For such labelled networks, we first derive a few enumerative results; we then discuss recursion relations and derive their generating function. Whenever suitable, the asymptotics will be discussed, and we will also give references to the Online Encyclopedia of Integer Sequences~\cite{oeis}, when the counting sequence in question can be found there.

 The combinatoric multiplicities which we will find can be used, for instance, to find all possible configurations of a chromatin segment with a given number of TUs and a list of desired features (such as the number of clusters and of singletons). This provides a useful bound for all possible topologies that this genomic region can form, in either simulations or experiments. 

\subsection{Configurations with an arbitrary number of clusters}

To begin with, we note that, if we do not care about the number of clusters in the configurations, then the number of different configurations with $n$ TUs  is given by the {\em Bell number} $B_n$. This is because each configuration can be thought of as a partition of the set $[n]$, where each subset (or block, or part) with at least two TUs will form a cluster, while the singletons will correspond to the TUs not belonging to any clusters. For example, for $n=6$, the partition $\{\{1,4\},\{2\},\{3,5,6\}\}$ encodes a possible configuration with two clusters.  It is well-known that $B_n$ counts the number of partitions of $[n]$, and this is the sequence A000110 in \cite{oeis} that begins with 
\begin{equation}
1, 2, 5, 15, 52, 20, 877, 4140, \ldots.
\end{equation}
The Bell numbers satisfy the recurrence relation $B_{n+1}=\sum_{k=0}^{n}{n\choose k}B_k$. Their {\em exponential generating function} $\sum_{n\geq 0}B_n\frac{t^n}{n!}$ is $e^{e^t-1}$, while the ordinary generating function is
\begin{equation}
B(t)=\sum_{k\geq 0}\frac{t^k}{\prod_{j=1}^{k}(1-jt)}.
\end{equation}
Also, the Bell numbers satisfy Dobinski's formula $B_n=\frac{1}{e}\sum_{k=0}^{\infty}\frac{k^n}{k!}$ and asymptotically ($n\to\infty$),
\begin{equation}
B_n\sim\frac{1}{\sqrt{n}}\left(\frac{n}{W(n)}\right)^{n+\frac{1}{2}}\exp\left(\frac{n}{W(n)}-n-1\right),
\end{equation}
where the {\em Lambert W function} has the same growth as the logarithm~\cite{Lovasz1993}.

Interestingly, if $B^*_n$ denotes the number of configurations {\em without singletons} (i.e.\ each TU is part of a cluster), then the following (well-known) combinatorial argument can be used to show that $B^*_{n+1}=B_n-B^*_n$. Note that $B_n-B^*_n$ is the number of partitions of $[n]$ that have at least one singleton. Now, take all singletons in a partition counted by $B_n-B^*_n$ and add them together along with the element $n+1$ so that to form a subset in a partition of $[n+1]$ that has no singletons (and hence is counted by $B^*_{n+1}$). This mapping is a bijection. 
 
\subsection{Configurations with a fixed number of clusters}

We now discuss how to enumerate configurations with a fixed number of clusters. To do so, a useful set of quantities is provided by the {\em Stirling numbers of the second kind} $S(n,k)$, which count the number of ways to partition the set $[n]$ into $k$ subsets. Even though $S(n,k)$ does not directly give us the number of configurations with $k$ clusters, below we will make use of these numbers.

We wish to find the number of partitions of $[n]$ into subsets (i.e., the number of configurations) so that precisely $k$ subsets, $1\leq k\leq n-2$, have two or more elements (i.e., with precisely $k$ clusters). We call this number $f(n,k)$. We highlight that this quantity counts the partition of $n$ TUs into $k$ clusters, with an arbitrary number of singletons. 

\subsubsection{Number of configurations with one cluster} 

There are $2^n-n-1$ configurations corresponding to the case of $k=1$. Indeed, each binary string $s_1s_2\ldots s_n$ over the alphabet $\{0,1\}$ corresponds to a configuration, where $s_i=0$ indicates that the TU $i$ is a singleton, while $s_i=1$ indicates that the TU $i$ is included in the only cluster. The number of possibilities is $2^n$, but we need to subtract the situations when at most one $1$ is present in the string because a cluster needs to have at least two TUs. Note that asymptotically, we have $O(2^n)$ such configurations. For $n\geq 1$, the counting sequence begins
\begin{equation}
0, 1, 4, 11, 26, 57, 120, 247, 502, \ldots
\end{equation}
and this is the sequence A000295 in \cite{oeis}.

\subsubsection{Number of configurations with two clusters} 

The case of $k=2$ can be derived similarly to the case of $k=1$. Instead of binary sequences, we can consider sequences over $\{0,1,2\}$ (there are $3^n$ of them), and then subtract those sequences that do not correspond to configurations with precisely two clusters (for example, sequences with no $2$s, or with one $1$ and one $2$). However, this method is still cumbersome, so we use the following formula instead, where $i$ corresponds to the number of singletons in a configuration (this number cannot be bigger than $n-4$ for us to be able to create two clusters), ${n\choose i}$ is the number of ways to choose these singletons in $[n]$, $S(n-i,2)$ (equal to $2^{n-i-1}-1$~\cite{Rennie1969}) counts the number of configurations with two clusters, and we have to subtract $(n-i)$, the number of possibilities for clusters receiving a single TU, giving 
\begin{eqnarray}
f(n,2)&=&\sum_{i=0}^{n-4}{n\choose i}(S(n-i,2)-(n-i))=\sum_{i=0}^{n-4}{n\choose i}(2^{n-i-1}-1-n+i)\\ \nonumber
&=&\frac{1}{2}(3^{n}+1)-(n+2)2^{n-1}+{n\choose 2}+n.
\end{eqnarray}
The last equality can be checked, for instance, by induction. %is obtain using Maple, but can also be obtained by hand (and checked by induction).  
We note that asymptotically, the number of configurations is $O(3^n)$ and the counting sequence begins, for $n\geq 4$, with
\begin{equation}
3, 25, 130, 546, 2037, 7071,\ldots;
\end{equation}
this is the sequence A112495 in \cite{oeis}.

\subsection{Recurrence relations and %exponential
generating function for loop networks with an arbitrary number of singletons}

Using the approaches above is rather cumbersome to produce explicit formulas for arbitrary $k$. %, although starting with sequences over $\{1,2,\ldots,k+1\}$ to encode configurations with $k$ clusters, one could conjecture (which is yet to be proved rigorously!) that asymptotically, the number of configurations is $O((k+1)^n)$ for $k=0,1,2,\ldots$ up to a certain number depending on $n$. However,
Alternatively, we can produce a recurrence relation for the numbers in question, $f(n,k)$, that can be turned into a partial differential equation for the respective generating function.

%Let $f(n,k)$ be the number of configurations with $n$ TUs and $k$ clusters. 
Note that for $n\geq 2$, this recursion reads as follows,
\begin{equation}\label{rec-rel-1}
f(n,k)=(k+1)f(n-1,k)+(n-1)f(n-2,k-1).
\end{equation}
To prove Eq.~(\ref{rec-rel-1}), we can think of producing, in a unique way, a configuration with $n$ TUs from a smaller configuration by introducing the $n$-th TU. The possible disjoint options are: 
\begin{enumerate}[label=(\roman*)]
\item $n$ joins an existing cluster (with at least two TUs in it) or $n$ becomes a singleton, and there are $(k+1)f(n-1,k)$ possibilities in this case;
\item $n$ ends up in a cluster with exactly two TUs, and there are $(n-1)f(n-2,k-1)$ ways as there are $n-1$ ways to select a TU to share the cluster with $n$. 
\end{enumerate}
The initial conditions of \eqref{rec-rel-1} are $f(0,0)=1$ and $f(0,k)=0$ for $k\neq 0$, and $f(1,0)=1$ and $f(1,k)=0$ for $k\neq 0$, along with $f(n,k)=0$ for $k<0$. 

%According to A124324 in \cite{oeis}, 
We now consider the exponential generating function, defined as 
\begin{equation}
F(t,x)=\sum_{n,k\geq 0}f(n,k)\frac{t^n}{n!}x^k.
\end{equation}
We can write that
\begin{eqnarray}
\frac{\partial}{\partial t} F(t,x)&=&\sum_{n\ge 1, k\geq 0} f(n,k)\frac{t^{n-1}}{(n-1)!} x^k \\ 
&=& \sum_{n \geq 1; k\geq 1} k f(n-1,k)\frac{t^{n-1}}{(n-1)!}x^k + \sum_{n \geq 1; k\geq 0} f(n-1,k)\frac{t^{n-1}}{(n-1)!}x^k + \sum_{n \geq 2; k\geq 1} f(n-2,k-1) \frac{t^{n-1}}{(n-2)!}x^k \nonumber \\ 
&=& x\sum_{n, k\geq 1}k f(n,k)\frac{t^{n}}{n!}x^{k-1} 
+ \sum_{n, k\geq 0} f(n,k)\frac{t^{n}}{n!}x^k 
+ t x \sum_{n, k\geq 0} f(n,k) \frac{t^{n}}{n!}x^k
\nonumber \\ 
&=& x \frac{\partial }{\partial x}F(t,x)+F(t,x)(1+tx). \end{eqnarray}
Therefore, $F(t,x)$ satisfies the following partial differential equation,
\begin{equation}
\frac{\partial}{\partial t}F(t,x) -x \frac{\partial }{\partial x}F(t,x)= (1+tx) F(t,x).
\end{equation}
The solution of this equation with the boundary conditions $F(0,x)=1$ and $F(t,0)=e^t$ can be found explicitly to be 
\begin{equation}\label{egff}
F(t,x) = e^{x e^t-x+(1-x)t}.
\end{equation}
Eq.~(\ref{egff}) can be used to find $f(n,k)$ for arbitrary values of $n$ and $k$, as well as the associated asymptotic behaviour. Note that $f(n,k)$ is the sequence known as A124324 in~\cite{oeis}, where Eq.~(\ref{egff}) is also given.

\if{
Another useful quantity is 
\begin{equation}
f_k(t)=\sum_{n\geq 0}f(n,k)\frac{t^n}{n!},
\end{equation}
which can be shown to obey the following differential equation,
\begin{equation}
\frac{d}{dt}f_k(t) = (k+1) f_k(t) + t f_{k-1}(t).
\end{equation}
Note that $f_{k}(t)=0$ for $k<0$, and that the boundary condition $f_k(0)=0$ needs to be satisfied. The equation is difficult to solve due to the condition that $f_{k}(t)=0$ for $k<0$. It is however possible to find an asymptotic solution for $k\to\infty$, which is given by
\begin{equation}\label{egff1}
f_k(t) = \frac{\left(e^t-1-t\right)^{k+1}}{\left(k+1\right)!}.
\end{equation}
}\fi

%\begin{equation}\label{pde2}
%\frac{\partial }{\partial x}G_0(t,x) + \frac{t^2}{x} \frac{\partial }{\partial t}G_0(t,x)=\frac{(1-t-t^2)G_0(t,x)}{tx}.
%\end{equation}

\subsection{Loop networks with a fixed number of singletons}

We can refine Eq.~(\ref{rec-rel-1}) to enumerate configurations with a fixed number of singletons. Let $f(n,k,\ell)$ be the number of configurations with $n$ TUs, $k$ clusters, and $\ell$ singletons. This quantity satisfies the following recursion relation:
\begin{equation}\label{recursionrelation2}
f(n,k,\ell)=kf(n-1,k,\ell)+f(n-1,k,\ell-1)+(\ell+1)f(n-1,k-1,\ell+1).
\end{equation}
Indeed, we can think of producing, in a unique way, a configuration with $n$ TUs from a configuration with $n-1$ TUs by introducing the TU $n$. The possible disjoint options are: 
\begin{enumerate}[label=(\roman*)]
\item $n$ joins an existing cluster (with at least two TUs in it), and there are $kf(n-1,k,\ell)$ possibilities in this case;
\item $n$ is a singleton, and there are $f(n-1,k,\ell-1)$ possibilities in this case;
\item $n$ forms a cluster with precisely one other TU, in which case there are $(\ell+1)f(n-1,k-1,\ell+1)$ possibilities. 
\end{enumerate}

By iterating the recursion relation~(\ref{recursionrelation2}) with the easily checkable base $f(1,0,1)=1$ and $f(1,k,\ell)=0$ otherwise, we obtain
\begin{enumerate}[label=(\roman*)]
\item $f(2,0,1)=0$, $f(2,0,2)=1$, $f(2,1,0)=1$, recovering the total number of configurations with two TUs, $B_2=2$;
\item $f(3,0,2)=0$, $f(3,0,3)=1$, $f(3,1,0)=1$, $f(3,1,1)=3$, recovering the total number of configurations with three TUs, $B_3=5$;
\item $f(4,0,4)=1$, $f(4,1,0)=1$, $f(4,1,1)=4$, $f(4,1,2)=6$, $f(4,2,0)=3$, recovering the total number of configurations with four TUs, $B_4=15$;
\item $f(5,0,5)=1$, $f(5,1,0)=1$, $f(5,1,1)=5$, $f(5,1,2)=10$, $f(5,1,3)=10$, $f(5,2,0)=10$, $f(5,2,1)=15$  recovering the total number of configurations with five TUs, $B_5=52$, and so on.
\end{enumerate}

By using a similar approach as in Section IIC, we can define the following exponential generating function,
\begin{equation}
F(t,x,y)=\sum_{n,k,\ell\geq 0}f(n,k,\ell)\frac{t^n}{n!} x^k y^{\ell},
\end{equation}
which obeys the following partial differential equation,
\begin{equation}
\frac{\partial}{\partial t}F(t,x,y) - x \frac{\partial}{\partial x}F(t,x,y) - x\frac{\partial}{\partial y}F(t,x,y)= yF(t,x,y). 
\end{equation}
Quite remarkably, the physically relevant solution of this more complex equation can also be found explicitly and is given by 
\begin{equation}\label{egff2}
F(t,x,y)=e^{yt}e^{x(e^t-1-t)}.  
\end{equation}
Note that this solution satisfies the following boundary conditions: (i) $F(0,x,y)=1$; (ii) $F(t,0,y)=e^{yt}$; (iii) $F(t,x,0)=e^{x(e^t-1-t)}$. Once more, Eq.~(\ref{egff2}) can be expanded to yield coefficients $f(n,k,\ell)$, therefore solving the problem of enumerating all configurations with a fixed number of TUs, clusters, and singletons.

\if{In analogy with Section IIC, we can additionally define the related quantities, 
\begin{equation}
f_{k,\ell}(t)=\sum_{n}f(n,k,\ell)\frac{t^n}{n!},
\end{equation}
which obey the set of differential equations
\begin{equation}
\frac{d}{dt}f_{k,\ell}(t) = k f_{k,\ell}(t) + f_{k,\ell-1}(t)+(l+1) f_{k-1,\ell+1}(t),
\end{equation}
subject to the boundary condition $f_{k,l}(0)=0$, and where $f_{k,l}(t)$ is implied to be $0$ when either $k$ or $l$ are negative. For $k$ and $\ell$ large, the asymptotic solution can be found explicitly to be
\begin{equation}\label{egff3}
f_{k,\ell}(t) = \frac{\left(e^t-1\right)^{k+\ell+1}}{\left(k+\ell+1\right)!}.
\end{equation}
}\fi

\if{We can also define the ordinary generating function as  $F(t,x,y)=\sum_{n\geq1}\sum_{k,\ell\geq 0}f(n,k,\ell)t^nx^ky^{\ell}$ and write
\begin{eqnarray}
F(t,x,y)&=&\sum_{n\geq1}\sum_{k,\ell\geq 0}f(n,k,\ell)t^nx^ky^{\ell} \\ 
&=& ty + \sum_{n\geq 2}\sum_{k,\ell\geq 0}kf(n-1,k,\ell)t^nx^ky^{\ell} \nonumber \\ 
&& \ \ \   + \sum_{n\geq 2}\sum_{k,\ell\geq 0}f(n-1,k,\ell-1)t^nx^ky^{\ell}\nonumber  \\ 
&& \ \ \   + \sum_{n\geq 2}\sum_{k,\ell\geq 0}(\ell+1)f(n-1,k-1,\ell+1)t^nx^ky^{\ell} \nonumber \\ 
&=& ty + tx\sum_{n\geq 1}\sum_{k,\ell\geq 0}kf(n,k,\ell)t^nx^{k-1}y^{\ell} \nonumber \\ 
&& \ \ \   + ty\sum_{n\geq 1}\sum_{k,\ell\geq 0}f(n,k,\ell-1)t^nx^ky^{\ell-1}\nonumber  \\ 
&& \ \ \   + tx\sum_{n\geq 1}\sum_{k,\ell\geq 0}(\ell+1)f(n,k-1,\ell+1)t^nx^{k-1}y^{\ell} \nonumber \\ 
&=& ty + tx \frac{\partial }{\partial x}F(t,x,y)+tyF(t,x,y)+tx\frac{\partial }{\partial y}F(t,x,y).\nonumber 
\end{eqnarray}

Therefore, we find that $F(t,x,y)$ satisfies the following partial differential equation:
\begin{equation}
\label{pdegf}
\frac{\partial }{\partial x}F(t,x,y) +  \frac{\partial }{\partial y}F(t,x,y)=\frac{(1-yt)F(t,x,y)-yt}{tx}.
\end{equation}

While this equation can be solved with the method of characteristics, we note its general solution is less directly helpful for our enumeration aims with respect to that of the partial exponential generating function in Eq.~(\ref{egff2}).}\fi

\subsection{Results for networks without singletons}

It is sometimes useful, or of interest, to consider the case where there are no singletons in the configuration. This is, for instance, the case that is considered in the companion paper~\cite{short}. If we denote by $N(n,k)$ the number of configurations with $n$ TUs, $k$ clusters and no singletons, such that $N(n,k)=f(n,k,0)$, we find that for $k=2$
\begin{equation}\label{2clusters_nosingleton}
N(n,2)=2^{n-1}-n-1=f(n-1,1)-1=f(n-1,1)-N(n-1,1).
\end{equation}
This equation can be derived by noting that the configurations of the chain can be constructed by assigning the first bead to cluster $0$, and computing the number of configurations of the rest of the TUs with a single cluster and an arbitrary number of singletons. The singletons are then put in the same cluster as the first bead. In this way, we obtain all configurations with $2$ clusters and no singletons once we subtract the single configuration which has no singletons in the rest of the chain (as this would lead to a configuration where the first TU is a singleton, which does not contribute to $N(n,2)$).

A similar argument leads to the general identity
\begin{equation}
N(n,k)=f(n-1,k-1)-N(n-1,k-1),
\end{equation}
linking the number of configurations with a given number of clusters with and without singletons. 

%\begin{equation}
%N(n,3)=N_0(n-1,2)-N(n-1,2)
%\end{equation}

The quantities $N(n,k)$ obey the following recursion relation~\cite{Bona2016,Beyene2021,Nabawanda2020}
\begin{equation}\label{recursionrelation3}
N(n,k)=k N(n-1,k)+ (n-1) N(n-2,k-1).
\end{equation}

Similarly to what done previously, starting from Eq.~(\ref{recursionrelation3}) we can find the following exponential generating function
for $N(n,k)$, 
\begin{equation}
G(t,x)=\sum_{n\geq 0}N(n,k)\frac{t^n}{n!}x^k,
\end{equation}
to be given by
\begin{equation}\label{egff4}
G(t,x) = e^{x(e^t-1-t)}.
\end{equation}
The related quantities
\begin{equation}
g_k(t)=\sum_{n\geq 0}N(n,k)\frac{t^n}{n!},
\end{equation}
can now be found exactly for each $k$, and are given by~\cite{Bona2016}
\begin{equation}\label{egff5}
g_{k}(t) = \frac{\left(e^t-1-t\right)^{k}}{k!}.
\end{equation}

%\begin{eqnarray}
%G(t,x) & = & \sum_{n\ge 2} \sum_{k\ge 1} N(n,k) t^n x^k \\ \nonumber
%& = & t^2 x + \sum_{n\ge 3} \sum_{k\ge 1} k N(n-1,k) t^n x^k + \sum_{n\ge 3} \sum_{k\ge 1} (n-1) N(n-2,k) t^n x^k \\ \nonumber
%& = & t^2 x + tx \sum_{n\ge 2} \sum_{k\ge 1} k N(n,k) t^n x^{k-1} +  t^2 \sum_{n\ge 1} \sum_{k\ge 1} (n+1) N(n,k) t^{n} x^{k} \\ \nonumber
%& = & t^2 x + tx \sum_{n\ge 2} \sum_{k\ge 1} k N(n,k) t^n x^{k-1} +  t^2 \sum_{n\ge 2} \sum_{k\ge 1} (n+1) N(n,k) t^{n} x^{k} \\ \nonumber
%& = & t^2 x + tx \frac{\partial}{\partial x} G(t,x) +  t^3 \frac{\partial}{\partial t}G(t,x) +t^2 G(t,x).
%\end{eqnarray}

%\begin{equation}
%x \frac{\partial}{\partial x} G(t,x) + t^2 \frac{\partial}{\partial t} G(t,x) = \frac{G(t,x)-t^2 x}{t}
%\end{equation}

\subsection{String of rosettes and reducible networks}

A natural question is whether a particular configuration can be broken up, or reduced, into a series of simpler configurations. To characterise such states, we call 
a configuration with $n$ TUs {\em irreducible} if it contains no cluster and only singletons, or it has $k$ clusters, one of which contains TU $n$, and it is not possible to separate the $k$ clusters into two groups by cutting a single polymer segment. 

\begin{figure}
\begin{center}
\begin{tikzpicture}[scale=1.00]
    \SetGraphUnit{3}
    \Vertex{A}
    \EA(A){B}
    \WE[unit=2](A){in}
    \EA[unit=2](B){out}
    \Edge(A)(B)
    \Edge(in)(A)
    \Edge(B)(out)
    \loopNarrowN{A}
    \loopNarrowS{A}
    \loopNarrowWSW{A}
    \loopNarrowNW{B}
    \loopNarrowSW{B}
    \loopNarrowSE{B}
    \loopLongNE{B}

    \EA[unit=4](B){C}
    \Edge(B)(C)
    \loopLongS{C}
    \EA(C){D}
    \Edge(C)(D)
    \loopNarrowN{D}
    \loopNarrowS{D}
    \EA[unit=4](D){out2}
    \Edge(D)(out2)    

    \draw [fill=red] (-1,0) circle (0.1cm);    
    \draw [fill=red] (0,0) circle (0.1cm);    
    \draw [fill=red] (1,0) circle (0.1cm);    
    \draw [fill=red] (2,0) circle (0.1cm);   
    \draw [fill=red] (3,0) circle (0.1cm);   
    \draw [fill=red] (4.45,1.45) circle (0.1cm);   
    \draw [fill=red] (6,0) circle (0.1cm); 
    \draw [fill=red] (7,0) circle (0.1cm);
    \draw [fill=red] (7,-2.05) circle (0.1cm);    
    \draw [fill=red] (9,0) circle (0.1cm);    
    \draw [fill=red] (10,0) circle (0.1cm); 
    \draw [fill=red] (11.25,0) circle (0.1cm);  
    \draw [fill=red] (12,0) circle (0.1cm);    
    \draw [fill=red] (13.5,0) circle (0.1cm);  

    \draw (0.5,-2) node {} -- (0.5,2) [dashed] node {}; 
    \draw (5,-2) node {} -- (5,2) [dashed] node {}; 
    \draw (7.5,-2) node {} -- (7.5,2) [dashed] node {}; 
    \draw (10.5,-2) node {} -- (10.5,2) [dashed] node {}; 

\end{tikzpicture}
\end{center}
\caption{Example of a reducible network, and of its decomposition into irreducible blocks (here separated by dashed vertical lines). The configuration shown is a string of rosettes.}\label{return-free-config}
\end{figure}
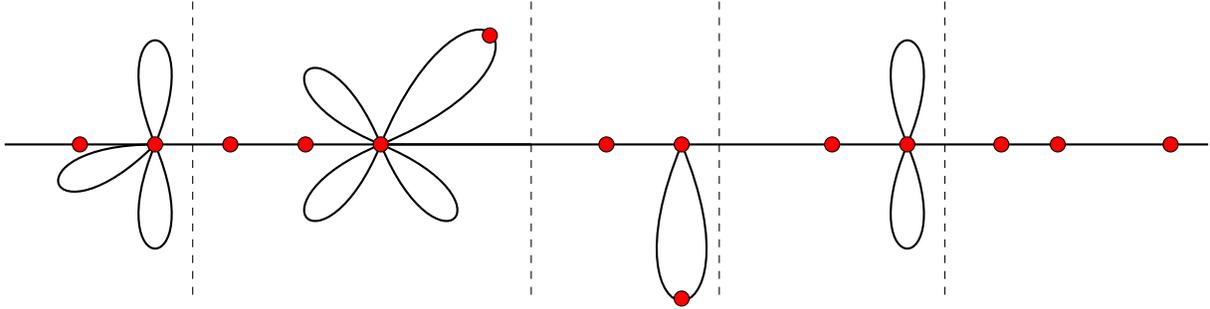

An example of a reducible network is a string of rosettes, shown in Figure~\ref{return-free-config}, where each irreducible component has a single cluster at most.  %({\bf This figure needs to be redrawn in black colour to match the style of the rest of the paper; words there, as well as brackets can be omitted, but vertical dashed lines should stay to show the decomposition}), can be decomposed into irreducible configurations, called {\em irreducible blocks}, as shown in Figure~\ref{return-free-config}. 
The decomposition into irreducible blocks is always unique assuming that if the configuration has at least one cluster, then the leftmost irreducible block has a cluster.  

We next derive the ordinary generating function $A(t)$ for the number of configurations in a string of rosettes. Note that there are $2^{n-1}-1$ irreducible configurations with $n$ TUs with a cluster as this is precisely the number of ways to choose at least one TU to join $n$ in the cluster. The generating function for these numbers is
\begin{eqnarray}I(t)&=&\sum_{n\geq 2}(2^{n-1}-1)t^n=t\sum_{n\geq 2}(2t)^{n-1}-\sum_{n\geq 2}t^n \\
&=& t\left(\frac{1}{1-2t}-1\right)-\left(\frac{1}{1-t}-t-1\right)=\frac{t^2}{(1-2t)(1-t)}.\nonumber
\end{eqnarray}
Noting that the generating function for irreducible blocks without clusters is $\frac{1}{1-t}$, we have
\begin{equation}
A(t)=\frac{1}{1-I(t)}\cdot\frac{1}{1-t}=\frac{1-2t}{1-3t+t^2}=1+t+2t^2+5t^3+13t^4+34t^5+89t^6+O(t^7).
\end{equation}
The corresponding sequence is A001519 in~\cite{oeis} and it has many combinatorial interpretations. One can derive from the generating function through the recurrence relation that %({\bf I haven't done it just taking the answer from \cite{oeis} but should be able to do it, if necessary}) 
\begin{equation}
    a_n = (\phi^{2n-1} + \phi^{1-2n})/\sqrt{5},
\end{equation}
where $\phi=(1+\sqrt{5})/2$, and hence asymptotically, the number of configurations in a string of rosettes is $O\left(\left(\frac{3+\sqrt{5}}{2}\right)^n\right)\approx O(2.618^n)$.

The concept of reducible networks would be useful to enumerate configurations of longer chains that we consider in this work. Additionally, strings of rosettes appear often in simulations and it is therefore useful to provide a way to separately count the number of possible configurations leading to this specific type of polymer network. 

\section{Combinatorics of inequivalent topologies for unlabelled networks}

We now discuss the case of unlabelled networks, which as anticipated is of interest when discussing the relative frequencies of different network topologies, irrespective of the specific labelling chosen. This is relevant, for instance, when asking whether, in a simulation, or experiment, rosette topologies are more or less common than watermelon ones. 

\begin{figure*}[h!]
	\begin{center}		
	\includegraphics[width=0.85\textwidth]{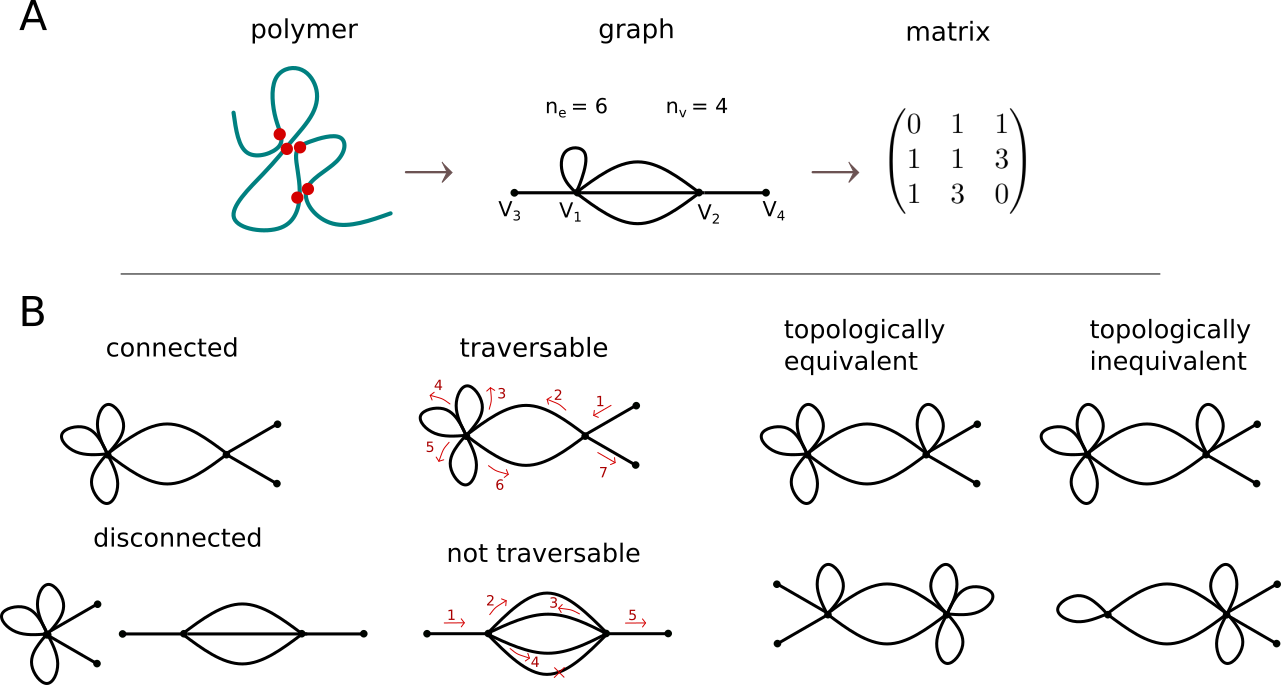}
\end{center}
\caption{\textbf{(A)} Schematics showing how polymer networks can be converted into graphs, and graphs to matrices. \textbf{(B)} Examples of connected and disconnected, traversable and not traversable, topologically equivalent and inequivalent graphs (or equivalently polymer networks).
	}
	\label{fig:graphs}
\end{figure*}

To study this case, we will be mapping networks to graphs and matrices. While this mapping is not necessary to derive the formula we will give below, which holds for $k=2$ clusters, it provides a useful framework to build, for instance, numerical algorithms which can enumerate all possible inequivalent topologies with a larger number of clusters $k$.

Specifically, we begin by noting that the network topologies assembled by joining the TUs of a polymer can be mapped to graphs with $n_v$ vertices and $n_e$ edges (see Fig.~\ref{fig:graphs}A). The vertices of the graph correspond to either cluster of TUs or to one of the two polymer ends, while the edges of the graph correspond to polymer segments between two TUs or between one TU and one of the polymer ends.

We note that not all graphs can be representations of a polymer with TUs: since they are associated with a folded polymer, graphs representative of a chromatin loop network must be connected and traversable (see Fig.~\ref{fig:graphs}B). Additionally, both $n_v$ and $n_e$ are constrained by the number of TUs $n$. If none of the TUs coincide with the ends of the polymer, and if singletons are disallowed (as in Fig.~\ref{fig:graphs}A), $n_e = n+1$ and $3<n_v\leq\lfloor\frac{2p+2}{3}\rfloor+2$, where $p=n_e-2$. Two of these vertices correspond to the polymer ends, and their degree is $1$; we will call \emph{internal} vertices all the others, namely all the vertices associated with clusters of TUs ($V_1$ and $V_2$ in Fig.~\ref{fig:graphs}A)~\footnote{The upper bond comes from requiring the degree of the internal vertices to be greater than or equal to 3. Since it is tied to traversability, the maximum number of vertices can be lower than this upper bound.}.

\subsection{Enumeration of inequivalent topologies with $2$ clusters}

We now proceed to count the number of topologically inequivalent, connected, and traversable graphs with a given number, $n+2$, of edges and $4$ vertices, $V_1$, $V_2$, $V_3$, $V_4$, two of which, $V_3$, $V_4$, of degree $1$. This is the number of topologically inequivalent networks with $n$ TUs and $k=2$ clusters, without any singletons, studied in the companion paper~\cite{short}.

Let $\mathcal{G}$ be a graph of this kind. $\mathcal{G}$ is identified by $5$ numbers: $a$, $b$, $c$, $n_1$ and $n_2$. Of these, $a$ and $b$ denote the number of edges connecting, respectively, vertex $V_1$ and vertex $V_2$ to themselves, $c$ is the number of edges connecting vertex $V_1$ to vertex $V_2$, while $n_1$ and $n_2$ are the numbers of vertices of degree $1$ connected, respectively, to $V_1$ and $V_2$ (for instance, the network in Fig.~\ref{fig:graphs}A has $n_1=n_2=1$, whereas the top left graph in Fig.~\ref{fig:graphs}B has $n_1=2$, $n_2=0$). The following symmetric matrix, therefore, identifies $\mathcal{G}$ in a compact way (see Fig.~\ref{fig:graphs}A)
\begin{equation}
	M(\mathcal{G})=
	\begin{pmatrix}
		0 & n_1 & n_2 \\
		n_1 & a & c \\
		n_2 & c & b 
	\end{pmatrix}.
\end{equation}

To count the number of graphs we are interested in, we remark that:
\begin{enumerate}[label=(\roman*)]
\item \label{it:ineq}
$\mathcal{G}$ and $\mathcal{G'}$ are equivalent if and only if 
\begin{equation}
	P^TM(\mathcal{G})P = M(\mathcal{G'}),
\end{equation}
where $P$ is one of the two permutation matrices
\begin{equation}
	\begin{pmatrix}
		1 & 0 & 0 \\
		0 & 1 & 0 \\
		0 & 0 & 1 
	\end{pmatrix}, \,
	\begin{pmatrix}
		1 & 0 & 0 \\
		0 & 0 & 1 \\
		0 & 1 & 0 
	\end{pmatrix}.
\end{equation}
In this case, we say that $M(\mathcal{G})$ and $M(\mathcal{G'})$ are equivalent (i.e., they represent equivalent graphs).

\item \label{it:conn}
$\mathcal{G}$ is disconnected if and only if $c = 0$.

\item \label{it:edges}
$n-1=a+b+c$.
\item \label{it:degs}
$\deg(V_1) = n_1+2a+c\geq3$ and $\deg(V_2) = n_2+2b+c\geq3$.

\item \label{it:transv}
Since a connected graph is traversable if and only if the number of vertices with odd degree is either $0$ or $2$~\cite{Trudeau2013}, $\deg(V_1)$ and $\deg(V_2)$ must be even. Moreover, since $\deg(V_1) = n_1+2a+c$ and $\deg(V_2) = n_2+2b+c$ we have the following cases: (A) if $c$ is even, either $n_1 = 2$ and $n_2 = 0$, or $n_1 = 0$ and $n_2 = 2$; (B)  if $c$ is odd, $n_1 = 1$ and $n_2 = 1$.

\end{enumerate}

Let us call $\{M\}_\mathcal{G}$ the set of all inequivalent (according to  point (i) above) matrices representing a graph with the desired constraints. To count the inequivalent topologies, let us consider the map $f: (a,b,c) \,|\, a,b,c\in\mathbb{N}, c\geq1, a+b+c =n \to M \in\{M\}_{\mathcal{G}}$ defined as
\begin{equation}
	f(a,b,c) =
	\begin{cases}
	\begin{pmatrix}
		0 & 2 & 0 \\
		2 & a & c \\	
		0 & c & b 
	\end{pmatrix}	\qquad & \text{if $c$ is even.}
	\vspace{0.1cm}\\
	\begin{pmatrix}
		0 & 1 & 1 \\
		1 & a & c \\	
		1 & c & b 
	\end{pmatrix} \qquad & \text{if $c$ is odd.}
	\end{cases}
\end{equation}

This map covers all the desired inequivalent topologies, but it is not injective~\footnote{Note that, for even $c$, we \emph{apparently} neglected $n_1=0, n_2=0$. This is because
\begin{equation*}
		\begin{pmatrix}
		1 & 0 & 0 \\
		0 & 0 & 1 \\
		0 & 1 & 0 
	\end{pmatrix}
		\begin{pmatrix}
		0 & 0 & 2 \\
		0 & b & c \\	
		2 & c & a 
	\end{pmatrix}
\begin{pmatrix}
	1 & 0 & 0 \\
	0 & 0 & 1 \\
	0 & 1 & 0 
\end{pmatrix} =
\begin{pmatrix}
	0 & 2 & 0 \\
	2 & a & c \\	
	0 & c & b 
\end{pmatrix}.
\end{equation*}}. The number of possible combinations of
$(a,b,c)$ satisfying the constraints $n=a+b+c$, $a\geq0$, $b\geq0$ and $c\geq1$ is given by
\begin{equation}
\sum_{i=1}^{n-1}(n-i)=\frac{n(n-1)}{2}.
\end{equation}
From this number, we first need to identify the  combinations of $(a,b,c)$ which map to equivalent graphs (or matrices), then to take away those which would lead to multiple counting of the same topology, and finally to remove the combinations which do not satisfy point~\ref{it:degs}. 

To do so, we note that the graph equivalence condition $P^TMP = M'$ with $M\ne M'$ requires $a=b'$, $b=a'$, $n_1=n_2'$ and $n_2=n_1'$. If $c$ is even, this is never met by construction; if $c$ is odd, $(a,b,c)$ and $(b,a,c)$ are mapped to equivalent matrices. To account for this, and avoid double counting of these equivalent topologies, we require $a\geq b$, a condition which removes $\sum_{odd\,c\le n}\frac{c-1}{2}$ possibilities: equivalently, $\sum_{i=1}^{\frac{n-1}{2}}i$ combinations if $n$ is odd, and $\sum_{i=1}^{\frac{n}{2}-1}i$ combinations if $n$ is even. 

Finally, to account for point~\ref{it:degs}, we also remove the two configurations $(a=n-2,b=0,c=2)$ and $(a=n-1,b=0,c=1)$ from the total count~\footnote{$(a=0,b=n-1,c=1)$ has already been removed, since $a<b$.}.

The total number of inequivalent graphs with $n$ TUs and $2$ clusters is then given by
\begin{equation}\label{nkeq2}
N_u(n,2)=\frac{n(n-1)}{2}-2-\sum_{i=1}^{\lfloor{\frac{n-1}{2}}\rfloor}i=\frac{n(n-1)}{2}-2-\frac{\lfloor{\frac{n-1}{2}\rfloor}\lfloor{\frac{n+1}{2}}\rfloor}{2},
\end{equation}
where $\lfloor x \rfloor$ denotes the floor of $x$ (the largest integral smaller than or equal to $x$).

%\subsection{Sticky ends}
\if{
For completeness, we note that the procedure explained above may be extended to count the number of inequivalent graph topologies in the singular case in which one or both of the polymer ends which are TUs, again without singletons.

If both ends are sticky, for instance, provided we remove $(n_e-1,0,1)$ and $(n_e-2,0,2)$ from the domain, the map $f: (a,b,c) \,|\, a,b,c\in\mathbb{N}, c\geq1, a+b+c =n_e, a\geq b \to M(\mathcal{G}) \in\{M\}_{\mathcal{G}}$ defined as
\begin{equation}
	f(a,b,c) =
		\begin{pmatrix}
			0 & 0 & 0 \\
			0 & a & c \\	
			0 & c & b 
		\end{pmatrix}
\end{equation}
is bijective. The total number of inequivalent graphs is then
\begin{equation}
\frac{n_e(n_e+1)}{2}-\sum_{i=0}^{n_e-1}\lceil\frac{i}{2}\rceil-2.
\end{equation}
}\fi

\subsection{Network multiplicities}

Note that, for each of the unlabelled network topologies  just found, there are multiple possible labelled configurations that correspond to it. As these combinatorial weights, or multiplicities, are generally different for different topologies, it is desirable to keep track of these. It is however difficult to go beyond a case-by-case study. Here, we focus on the case of $n=8$ TUs and $k=2$ clusters, studied in~\cite{short}, for which there are $20$ inequivalent topologies [as predicted by Eq.~(\ref{nkeq2}) for $n=8$]. 

In this case, for each inequivalent topology, Table~\ref{topology-table} provides the number of ties, $n_t$, the number of loops $n_l$, the degree of the two vertices in the graphs (corresponding to the clusters), $L_1$ and $L_2$ respectively, and the multiplicity of the topology $\Omega$, which is the number of labelled configurations corresponding to that topology. The degrees $L_1$ and $L_2$ determine the entropic exponent of the polymer network~\cite{Duplantier1989}: it can be seen that there are three classes of such exponents in the $20$ topologies considered in Table~\ref{topology-table}. 

Regarding multiplicities, we observe that these tend to be larger for hybrid networks which are intermediate between rosettes and watermelons and that multiplicities are in general small for networks with low $n_l$. This would suggest, that, in the absence of other biases, such configurations would form less often. We shall see in what follows, however, that these topologies are actually easier to form, so there is an interesting competition between the combinatoric multiplicity and the entropic cost of formation of these loop networks.

% topology-table
\newcommand{\TTA}{{Diagram}}
\newcommand{\TTC}{{$n_t$}}
\newcommand{\TTD}{{$n_l$}}
\newcommand{\TTE}{{$\Omega$}}
\newcommand{\TTK}{$\bert$}
\newcommand{\TTL}{{$L_1$}}
\newcommand{\TTM}{{$L_2$}}
\newcommand{\TTEN}{$\epsilon_c/(k_BT)$}
\newcommand{\TTCI}{\multicolumn{2}{c}{$\text{CI}/(k_BT)$}}
\begin{table}[!htbp]
\begin{center}
\sisetup{
    table-auto-round
}
\begin{tabular}{
    @{}
    l%                                                                          A diagrams
    S[table-format = 1]%                                                        C number of ties
    S[table-format = 1]%                                                        D number of loops
    S[table-format = 1]%                                                        L L_1
    S[table-format = 1]%                                                        M L_2
    %c%                                                                          K symmetry breaking
    S[table-format = 2]%                                                        E multiplicity for distinguishable sites
    c%S[table-format = 2.1]
    %S[table-format = 2.1]%                                                      G transition affinity
    >{{[}} %                                                                    H Add square bracket before column tex.stackexchange.com/questions/291786/
    S[table-format = 2.1,table-space-text-pre={[}]%                             H lower bound on confidence interval
    @{,\,} %                                                                    H Add comma and thin-space between the columns
    S[table-format = 2.1,table-space-text-post={]}]%                            H upper bound on confidence interval
    <{{]}} %   
    }
\toprule
%%%%%%%%%%%%%%%%%%%%%%%%%%%%%%%%%%%%%%%%%%%%%%%%%%%%%%%%%%%%%%%%%%%%%%%%%%
\TTA & \TTC & \TTD & \TTL & \TTM & \TTE & \TTEN & \TTCI \\ 
\colrule
%%%%%%%%%%%%%%%%%%%%%%%%%%%%%%%%%%%%%%%%%%%%%%%%%%%%%%%%%%%%%%%%%%%%%%%%%%
%\multirow{7}{*}[-23pt]{$\duplA$}% 162.83 / 2 = 81.415 -> -23pt adjustment
 \s\raisebox{-\totalheight/2}{\begin{tikzpicture}[scale=0.25,trim left]
    \SetGraphUnit{3}
    \Vertex{A}
    \EA(A){B}
    \NOWE[unit=1](A){in}
    \SOEA[unit=1](B){out}
    \Edge(A)(B)
    \Edge(in)(A)
    \Edge(B)(out)
    \loopNarrowN{A}
    \loopNarrowS{A}
    \loopNarrowWSW{A}
    \loopNarrowN{B}
    \loopNarrowS{B}
    \loopNarrowENE{B}
\end{tikzpicture}} &  1  &  6   &  8   &  8   &       1   &  9.1 & 9.0 & 9.5 \\
 \s\raisebox{-\totalheight/2}{\begin{tikzpicture}[scale=0.25,trim left]
    \SetGraphUnit{3}
    \Vertex{A}
    \EA(A){B}
    \NOWE[unit=1](A){in}
    \SOWE[unit=1](A){out}
    \Edge(in)(A)
    \Edge(A)(out)
    \loopNarrowN{A}
    \loopNarrowS{A}
    \loopNarrowE{B}
    \loopNarrowN{B}
    \loopNarrowS{B}
    \foreach \angle in {45}
        {
        \tikzset{EdgeStyle/.append style = {bend left = \angle}}
        \Edge(A)(B)
        \Edge(B)(A)
        };
\end{tikzpicture}} &  2   &  5   &  8   &  8   &       3   &  9.7 &  9.5 & 10.1 \\
 \s\raisebox{-\totalheight/2}{\begin{tikzpicture}[scale=0.25,trim left]
    \SetGraphUnit{3}
    \Vertex{A}
    \EA(A){B}
    \WE[unit=1](A){in}
    \EA[unit=1](B){out}
    \Edge(A)(B)
    \Edge(in)(A)
    \Edge(B)(out)
    \loopNarrowN{A}
    \loopNarrowS{A}
    \loopNarrowN{B}
    \loopNarrowS{B}
    \foreach \angle in {45}
        {
        \tikzset{EdgeStyle/.append style = {bend left = \angle}}
        \Edge(A)(B)
        \Edge(B)(A)
        };
\end{tikzpicture}} &  3   &  4   &  8   &  8   &      9    &  9.8 & 9.5 & 10.2 \\
 \s\raisebox{-\totalheight/2}{\begin{tikzpicture}[scale=0.25,trim left]
    \SetGraphUnit{3}
    \Vertex{A}
    \EA(A){B}
    \NOWE[unit=1](A){in}
    \SOWE[unit=1](A){out}
    \Edge(in)(A)
    \Edge(A)(out)
    \loopNarrowW{A}
    \loopNarrowENE{B}
    \loopNarrowESE{B}
    \foreach \angle in {22.5,45}
        {
        \tikzset{EdgeStyle/.append style = {bend left = \angle}}
        \Edge(A)(B)
        \Edge(B)(A)
        };
\end{tikzpicture}} &  4   &  3   &  8   &  8   &       9   & 10.4 & 10.1 & 10.5 \\
 \s\raisebox{-\totalheight/2}{\begin{tikzpicture}[scale=0.25,trim left]
    \SetGraphUnit{3}
    \Vertex{A}
    \EA(A){B}
    \NOWE[unit=1](A){in}
    \SOEA[unit=1](B){out}
    \Edge(A)(B)
    \Edge(in)(A)
    \Edge(B)(out)
    \loopNarrowWSW{A}
    \loopNarrowENE{B}
    \foreach \angle in {22.5,45}
        {
        \tikzset{EdgeStyle/.append style = {bend left = \angle}}
        \Edge(A)(B)
        \Edge(B)(A)
        };
\end{tikzpicture}} &  5   &  2   &  8   &  8   &      9    & 11.1 & 10.0 & 11.1 \\
 \s\raisebox{-\totalheight/2}{\begin{tikzpicture}[scale=0.25,trim left]
    \SetGraphUnit{3}
    \Vertex{A}
    \EA(A){B}
    \NOWE[unit=1](A){in}
    \SOWE[unit=1](A){out}
    \Edge(in)(A)
    \Edge(A)(out)
    \loopNarrowE{B}
    \foreach \angle in {15,30,45}
        {
        \tikzset{EdgeStyle/.append style = {bend left = \angle}}
        \Edge(A)(B)
        \Edge(B)(A)
        };
\end{tikzpicture}} &  6   &  1   &  8   &  8   &      3    & 10.9 & 10.6 & 11.5 \\
 \s\raisebox{-\totalheight/2}{\begin{tikzpicture}[scale=0.25,trim left]
    \SetGraphUnit{3}
    \Vertex{A}
    \EA(A){B}
    \WE[unit=1](A){in}
    \EA[unit=1](B){out}
    \Edge(A)(B)
    \Edge(in)(A)
    \Edge(B)(out)
    \foreach \angle in {15,30,45}
        {
        \tikzset{EdgeStyle/.append style = {bend left = \angle}}
        \Edge(A)(B)
        \Edge(B)(A)
        };
\end{tikzpicture}} &  7   &  0   &  8   &  8   &      1   & 11.2 & 11.0 & 11.4 \\ \colrule
%%%%%%%%%%%%%%%%%%%%%%%%%%%%%%%%%%%%%%%%%%%%%%%%%%%%%%%%%%%%%%%%%%%%%%%%%%
%\multirow{8}{*}[-45pt]{$\duplB$}% 215.5 pt / 2 = 107.75 -> -45pt adjustment
 \s\raisebox{-\totalheight/2}{\begin{tikzpicture}[scale=0.25,trim left]
    \SetGraphUnit{3}
    \Vertex{A}
    \EA(A){B}
    \WE[unit=1](A){in}
    \EA[unit=1](B){out}
    \Edge(A)(B)
    \Edge(in)(A)
    \Edge(B)(out)
    \loopNarrowN{A}
    \loopNarrowS{A}
    \loopNarrowNE{B}
    \loopNarrowNW{B}
    \loopNarrowSW{B}
    \loopNarrowSE{B}
\end{tikzpicture}} &  1   &  6   &  6   &  10   &  2  & 8.8 &  8.7 &  9.3  \\
 \s\raisebox{-\totalheight/2}{\begin{tikzpicture}[scale=0.25,trim left]
    \SetGraphUnit{3}
    \Vertex{A}
    \EA(A){B}
    \NOWE[unit=1](A){in}
    \SOWE[unit=1](A){out}
    \Edge(in)(A)
    \Edge(A)(out)
    \loopNarrowN{A}
    \loopNarrowS{A}
    \loopNarrowW{A}
    \loopNarrowENE{B}
    \loopNarrowESE{B}
    \foreach \angle in {45}
        {
        \tikzset{EdgeStyle/.append style = {bend left = \angle}}
        \Edge(A)(B)
        \Edge(B)(A)
        };
\end{tikzpicture}} &  2   &  5   &  10    & 6   &  4  &  9.6 &  9.4 & 10.0  \\
 \s\raisebox{-\totalheight/2}{\begin{tikzpicture}[scale=0.25,trim left]
    \SetGraphUnit{3}
    \Vertex{A}
    \EA(A){B}
    \NOWE[unit=1](A){in}
    \SOWE[unit=1](A){out}
    \Edge(in)(A)
    \Edge(A)(out)
%    \loopNarrowN{A}
%    \loopNarrowS{A}
    \loopNarrowW{A}
    \loopNarrowE{B}
    \loopNarrowN{B}
    \loopNarrowW{B}
    \loopNarrowS{B}
    \foreach \angle in {45}
        {
        \tikzset{EdgeStyle/.append style = {bend left = \angle}}
        \Edge(A)(B)
        \Edge(B)(A)
        };
\end{tikzpicture}} &  2   &  5   &  6   &  10   &  2  &  9.0 &  8.8 &  9.3  \\
 \s\raisebox{-\totalheight/2}{\begin{tikzpicture}[scale=0.25,trim left]
    \SetGraphUnit{3}
    \Vertex{A}
    \EA(A){B}
    \NOWE[unit=1](A){in}
    \SOEA[unit=1](B){out}
    \Edge(A)(B)
    \Edge(in)(A)
    \Edge(B)(out)
    \loopNarrowWSW{A}
    \loopNarrowWSW{A}
    \loopNarrowN{B}
    \loopNarrowS{B}
    \loopNarrowENE{B}
    \foreach \angle in {45}
        {
        \tikzset{EdgeStyle/.append style = {bend left = \angle}}
        \Edge(A)(B)
        \Edge(B)(A)
        };
\end{tikzpicture}} &  3   &  4   &  6   &  10   & 16 &  9.3 &  8.9 &  9.4   \\
 \s\raisebox{-\totalheight/2}{\begin{tikzpicture}[scale=0.25,trim left]
    \SetGraphUnit{3}
    \Vertex{A}
    \EA(A){B}
    \NOWE[unit=1](A){in}
    \SOWE[unit=1](A){out}
    \Edge(in)(A)
    \Edge(A)(out)
    \loopNarrowN{A}
    \loopNarrowS{A}
    \loopNarrowE{B}
    \foreach \angle in {22.5,45}
        {
        \tikzset{EdgeStyle/.append style = {bend left = \angle}}
        \Edge(A)(B)
        \Edge(B)(A)
        };
\end{tikzpicture}} &  4   &  3   &  10    & 6   & 12 & 10.1 & 10.0 & 11.0   \\
 \s\raisebox{-\totalheight/2}{\begin{tikzpicture}[scale=0.25,trim left]
    \SetGraphUnit{3}
    \Vertex{A}
    \EA(A){B}
    \NOWE[unit=1](A){in}
    \SOWE[unit=1](A){out}
    \Edge(in)(A)
    \Edge(A)(out)
    \loopNarrowN{B}
    \loopNarrowE{B}
    \loopNarrowS{B}
    \foreach \angle in {22.5,45}
        {
        \tikzset{EdgeStyle/.append style = {bend left = \angle}}
        \Edge(A)(B)
        \Edge(B)(A)
        };
\end{tikzpicture}} &  4   &  3   &  6   &  10   & 4  &  9.2 &  9.2 &  9.5   \\
 \s\raisebox{-\totalheight/2}{\begin{tikzpicture}[scale=0.25,trim left]
    \SetGraphUnit{3}
    \Vertex{A}
    \EA(A){B}
    \WE[unit=1](A){in}
    \EA[unit=1](B){out}
    \Edge(A)(B)
    \Edge(in)(A)
    \Edge(B)(out)
    \loopNarrowN{B}
    \loopNarrowS{B}
    \foreach \angle in {22.5,45}
        {
        \tikzset{EdgeStyle/.append style = {bend left = \angle}}
        \Edge(A)(B)
        \Edge(B)(A)
        };
\end{tikzpicture}} &  5   &  2   &  6   &  10   & 12 &  9.8 &  9.5 & 10.2   \\
 \s\raisebox{-\totalheight/2}{\begin{tikzpicture}[scale=0.25,trim left]
    \SetGraphUnit{3}
    \Vertex{A}
    \EA(A){B}
    \NOWE[unit=1](A){in}
    \SOWE[unit=1](A){out}
    \Edge(in)(A)
    \Edge(A)(out)
    \loopNarrowW{A}
    \foreach \angle in {15,30,45}
        {
        \tikzset{EdgeStyle/.append style = {bend left = \angle}}
        \Edge(A)(B)
        \Edge(B)(A)
        };
\end{tikzpicture}} &  6   &  1   &  10    & 6   & 4  & 10.6 & 10.5 & 11.3  \\  \colrule
%%%%%%%%%%%%%%%%%%%%%%%%%%%%%%%%%%%%%%%%%%%%%%%%%%%%%%%%%%%%%%%%%%%%%%%%%%
%\multirow{5}{*}[-27pt]{$\duplC$}% 143.8 / 2 = 81.416407075 -> -27pt adjustment
 \s\raisebox{-\totalheight/2}{\begin{tikzpicture}[scale=0.25,trim left]
    \tikzmath{
        \angleA = 30;
        \angleB = 70;
        \angleC = 110;
        \angleD = 150;
        \angleE = 210;
        \angleF = 270;
        \angleG = 330;
        }
    \SetGraphUnit{3}
    \Vertex{A}
    \EA(A){B}
    \NOWE[unit=1](A){in}
    \EA[unit=1](B){out}
    \Edge(A)(B)
    \Edge(in)(A)
    \Edge(B)(out)
    \loopNarrowWSW{A}
%    \loopNarrowENE{B}
%    \loopNarrowESE{B}
%    \loopNarrowN{B}
%    \loopNarrowS{B}
%    \loopNarrowWSW{B}
%    \foreach \angle in {45}
%        {
%        \tikzset{EdgeStyle/.append style = {bend left = \angle}}
%        \Edge(A)(B)
%        \Edge(out)(B)
%        % \Edge(B)(A)
%        };
    \Loop[
        dist = 2cm,
        dir = NO,
        style={
            out=\angleA,
            in=\angleB
            }
        ](B)%
    \Loop[
        dist = 2cm,
        dir = NO,
        style={
            out=\angleB,
            in=\angleC
            }
        ](B)%
    \Loop[
        dist = 2cm,
        dir = NO,
        style={
            out=\angleC,
            in=\angleD
            }
        ](B)%
    \Loop[
        dist = 2cm,
        dir = NO,
        style={
            out=\angleE,
            in=\angleF
            }
        ](B)%
    \Loop[
        dist = 2cm,
        dir = NO,
        style={
            out=\angleF,
            in=\angleG
            }
        ](B)%
\end{tikzpicture}} &  1   &  6   &  4   &  12   & 2 &  8.7 &  8.6 &  9.0   \\
 \s\raisebox{-\totalheight/2}{\begin{tikzpicture}[scale=0.25,trim left]
    \SetGraphUnit{3}
    \Vertex{A}
    \EA(A){B}
    \NOWE[unit=1](A){in}
    \SOWE[unit=1](A){out}
    \Edge(in)(A)
    \Edge(A)(out)
    \loopNarrowE{A}
    \loopNarrowN{A}
    \loopNarrowW{A}
    \loopNarrowS{A}
    \loopNarrowE{B}
    \foreach \angle in {45}
        {
        \tikzset{EdgeStyle/.append style = {bend left = \angle}}
        \Edge(A)(B)
        \Edge(B)(A)
        };
\end{tikzpicture}} &  2   &  5   &  12    & 4   & 5 &  9.1 &  8.8 &  9.3   \\
 \s\raisebox{-\totalheight/2}{\begin{tikzpicture}[scale=0.25,trim left]
    \SetGraphUnit{3}
    \Vertex{A}
    \EA(A){B}
    \NOWE[unit=1](A){in}
    \SOWE[unit=1](A){out}
    \Edge(in)(A)
    \Edge(A)(out)
    \loopNarrowENE{B}
    \loopNarrowN{B}
    \loopNarrowW{B}
    \loopNarrowS{B}
    \loopNarrowESE{B}
    \foreach \angle in {45}
        {
        \tikzset{EdgeStyle/.append style = {bend left = \angle}}
        \Edge(A)(B)
        \Edge(B)(A)
        };
\end{tikzpicture}} &  2   &  5   &  4   &  12   & 1 &  8.8 &  8.5 &  9.0   \\
 \s\raisebox{-\totalheight/2}{\begin{tikzpicture}[scale=0.25,trim left]
    \SetGraphUnit{3}
    \Vertex{A}
    \EA(A){B}
    \WE[unit=1](A){in}
    \EA[unit=1](B){out}
    \Edge(A)(B)
    \Edge(in)(A)
    \Edge(B)(out)
    \loopNarrowNE{B}
    \loopNarrowN{B}
    \loopNarrowS{B}
    \loopNarrowSE{B}
    \foreach \angle in {45}
        {
        \tikzset{EdgeStyle/.append style = {bend left = \angle}}
        \Edge(A)(B)
        \Edge(B)(A)
        };
\end{tikzpicture}} &  3   &  4   &  4   &  12   & 10 &  9.0 &  8.5 &  9.3  \\
 \s\raisebox{-\totalheight/2}{\begin{tikzpicture}[scale=0.25,trim left]
    \SetGraphUnit{3}
    \Vertex{A}
    \EA(A){B}
    \NOWE[unit=1](A){in}
    \SOWE[unit=1](A){out}
    \Edge(in)(A)
    \Edge(A)(out)
    \loopNarrowN{A}
    \loopNarrowW{A}
    \loopNarrowS{A}
    \foreach \angle in {22.5,45}
        {
        \tikzset{EdgeStyle/.append style = {bend left = \angle}}
        \Edge(A)(B)
        \Edge(B)(A)
        };
\end{tikzpicture}} &  4   &  3   &  12    & 4   & 10 &  9.3 &  9.1 &  9.3  \\\botrule
%%%%%%%%%%%%%%%%%%%%%%%%%%%%%%%%%%%%%%%%%%%%%%%%%%%%%%%%%%%%%%%%%%%%%%%%%%
\end{tabular}
\end{center}
\caption[Topology Summary Table]{Topology summary table. All topologies with $n = 8$ binding sites and $2$ clusters (graph vertices) are listed, together with their number of ties ($n_t$), number of loops ($n_t$), nontrivial vertex orders ($L_1$ and $L_2$ for first and second cluster), and multiplicity ($\Omega$). The last two columns give the critical energy between TUs needed to form the topology, $\epsilon_c$, in units of $k_BT$, together with the $95\%$ confidence interval: these results correspond to the simulations presented in Section IV. We have subdivided the diagrams into three classes, each characterised by the same pair of nontrivial vertex orders.}
\label{topology-table}
\end{table}

Note that, as required, the total sum of multiplicities for all topologies equals $N(8,2)=119$, namely the  number of $2-$cluster configurations with $n=8$ without singletons  found previously (see \eqref{2clusters_nosingleton}).

\section{Coarse-grained molecular dynamics simulations}

Having discussed the combinatorial problem of enumerating the possible configuration and inequivalent topologies of a chromatin loop network, we now turn to the associated polymer physics problem and ask what interaction between TUs needs to be included to form a target topology in practice. This calculation requires computer simulations for polymer models representing chromatin fibres, and therefore here we use coarse-grained molecular dynamics simulations to study this problem.

In this Section, we will first describe the model used, and then present our simulation results, whose main outcome will be to show that different topologies require significantly different energy inputs to form. This energy of formation will combine in practical examples with the combinatoric multiplicities discussed above (the relevant ones for the case at hand are those given in Table~\ref{topology-table}) to determine the likeliness of observing a given topology in an unconstrained polymer simulation, such as the one discussed in~\cite{short}.

\subsection{Model and potentials used}

We model a chromatin fibre as a bead-and-spring polymer. The underlying equations of motion are the set of Langevin equations for each bead, 
\begin{equation}
	m \hderiv{2}{\ve{x}_{i}}{t}
%	&
    = -\ve{\nabla}_{i} U - \gamma \deriv{\ve{x}_{i}}{t} + \sqrt{2 k_{B} T \gamma} \ve{\eta}(t)
    ,
\end{equation}
where $m$ is the bead mass, $\ve{x}_{i}$ is the $i$\nth bead position, $\gamma$ is the drag, and $\ve{\eta}(t)$ is uncorrelated white noise defined by $\av{\ve{\eta}(t)} = \ve{0}$ and $\av{\eta_{\alpha}(t) \eta_{\beta}(t^{\prime})} = \delta_{\alpha \beta} \delta(t-t^{\prime})$. This Langevin equation imposes an NVT ensemble on the system within which fluctuation and dissipation govern the exploration of the configuration space. 

In order to reproduce behaviour appropriate to chromatin, the following potentials, $U$, enter into this equation according to the particular beads under consideration.
% The different potentials enter into this equation to simulate the appropriate behaviour of chromatin.
A simple phenomenological Lennard--Jones potential, truncated to include only the repulsive regime (Weeks--Chandler--Andersen potential) acts between all beads in the system enforcing excluded volume, or self-avoidance. This is given by
\begin{equation}
	U_{\text{LJ}} (r_{ij}) =
%	&=
	\begin{cases}
		4 \varepsilon\left[\left(\frac{\sigma}{r_{i j}}\right)^{12}-\left(\frac{\sigma}{r_{i j}}\right)^{6}\right]+\varepsilon	&	\text{if } r_{i j}<2^{1 / 6} \sigma	\\
		0		&	\text{otherwise,}
	\end{cases}
\end{equation}
where $\sigma$ is the bead diameter. 

To capture chain connectivity, finitely extensible non-linear elastic (FENE) bonds are considered, acting only between consecutive beads along the polymer chain:
\begin{equation}
	U_{\text{FENE}} (r)
        %	&
        =-\tfrac12 K R_0^2 \ln \left[1-\left(\frac{r}{R_{0}}\right)^{2}\right]
    ,
\end{equation}
where $K = \SI{30}{\kT}/\sigma^2$ is the spring constant and $R_0=1.6\sigma$ is the maximum extent of the bond. 

Finally, we added  a bending or Kratky--Porod potential, which acts on the angle $\theta$ between three consecutive beads along  the chain and enforces a non-zero persistence length, $l_p$: 
\begin{equation}
	U_{\text{bending}}
        %	&
        =K_{b}[1+\cos (\theta)]
    ,
\end{equation}
where $K_{b} = k_B T l_p / \sigma = \SI{3}{\kT}$. Note that the persistence length is artificially raised during equilibration to assist the system in reaching a self-avoiding configuration. It is then lowered from $10\sigma$ to $3\sigma$ during the main simulation. This is an appropriate value for flexible chromatin~\cite{Brackley2013,Brackley2016,Brackley2021}.

To study the formation of a target topology we  included attractive interactions between beads which should be in the same cluster in the target topology; this procedure is  similar to what is done in a Go model approach to study protein folding, where only attractive interactions between residues in contact in the folded state are included~\cite{Onuchic2004}. 
We considered all $20$ topologies in Table~\ref{topology-table}; for instance, for a symmetric $2$-rosette state we included an interaction between the first four $TU$s, and between the last four. The attraction between the selected TUs was simulated by a Lennard-Jones potential, where part of the attractive tail was retained. The interaction range (cut-off) was set to $1.8 \sigma$, while the interaction strength $\epsilon$ was adjusted between $5$ and \SI{15}{\kT}. For large enough $\epsilon$, the target topology is formed, with all interactions realised. [Note that not all topology may be realisable if the steric interactions between different polymer segments prevent this, but in our case, this was not an issue.]

The transition between an initially unstructured chain and the target topology arises because the energy gained as binding sites come together increases (asymptotically) linearly with the number of binding beads in a cluster, whereas the entropic cost of adding legs to a cluster scales superlinearly~\cite{Duplantier1986,Duplantier1989,Marenduzzo2009}. As such, there is a critical energy $\epsilon_c$ at which the energy gain just offsets the entropic loss, and this is the transition point. Our goal was to find how the value of $\epsilon_c$ depends on topology. Note that all topologies we compare (i.e., within each of the three classes in Table~\ref{topology-table}) contain the same number of binding site interactions, hence the total maximum energy. The difference in $\epsilon_c$ is then primarily due to the entropic cost of forming that specific target topology. 

Our script loaded a modular pair coefficient file generated by a simple Python script. This allowed the target polymer network topologies to be easily specified and changed while keeping other elements of the simulation fixed, which was important for reproducibility, scalability, and comparisons. 
Finally, the system was evolved via Langevin dynamics  using the {\small LAMMPS} simulation package~\cite{plimpton:1995}.

\subsection{Target topology simulations: rosettes, watermelons, and dependence on the number of ties}

As discussed above, in the thermodynamic limit it is expected that the entropic exponent of forming a given topology should  depend only on the number ($L_{1,2}$) of legs meeting at its vertices~\cite{Duplantier1986,Duplantier1989}. This partitions the set of $20$ inequivalent topologies into three classes that have the same values of $L_{1,2}$, (see Table~\ref{topology-table}) so that results discussed below should be compared only between topologies in the same class. 

\begin{figure}
\begin{center}
\includegraphics{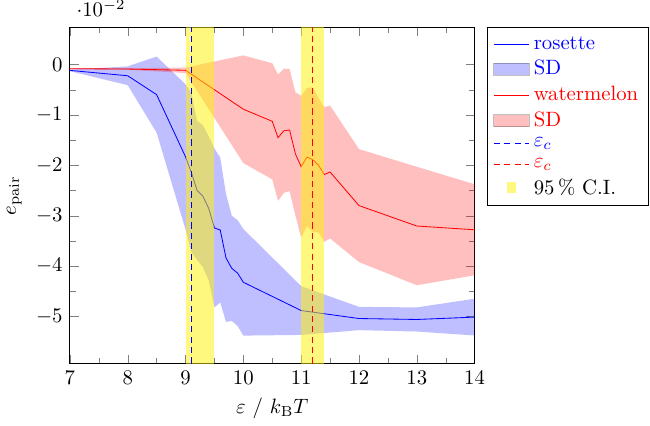}
\end{center}
\caption{Plot of the average energy per bead as a function of attractive energy $\epsilon$. The sigmoidal shape of the curve signals the formation of the target topology.}
\label{r-w-transition-plot}
\end{figure}

Amongst the first class (first seven topologies in Table~\ref{topology-table}), two topologies, namely the \emph{rosette} (top topology in the class) and the \emph{watermelon} (bottom topology in the class), stand out as particularly illustrative choices to discuss the results of the simulations. For these two  topologies, simulations were carried out by varying the interaction affinity $\varepsilon$ between $5 k_BT$ and $15 k_BT$. From the estimate of the
pairing energy ($\epair$) we computed the dimensionless quantity, $\epairNorm = \epair / \varepsilon$ that is reported in  figure~\ref{r-w-transition-plot} as a function of $\varepsilon$.

\begin{figure}
\begin{center}
\includegraphics[scale=0.25]{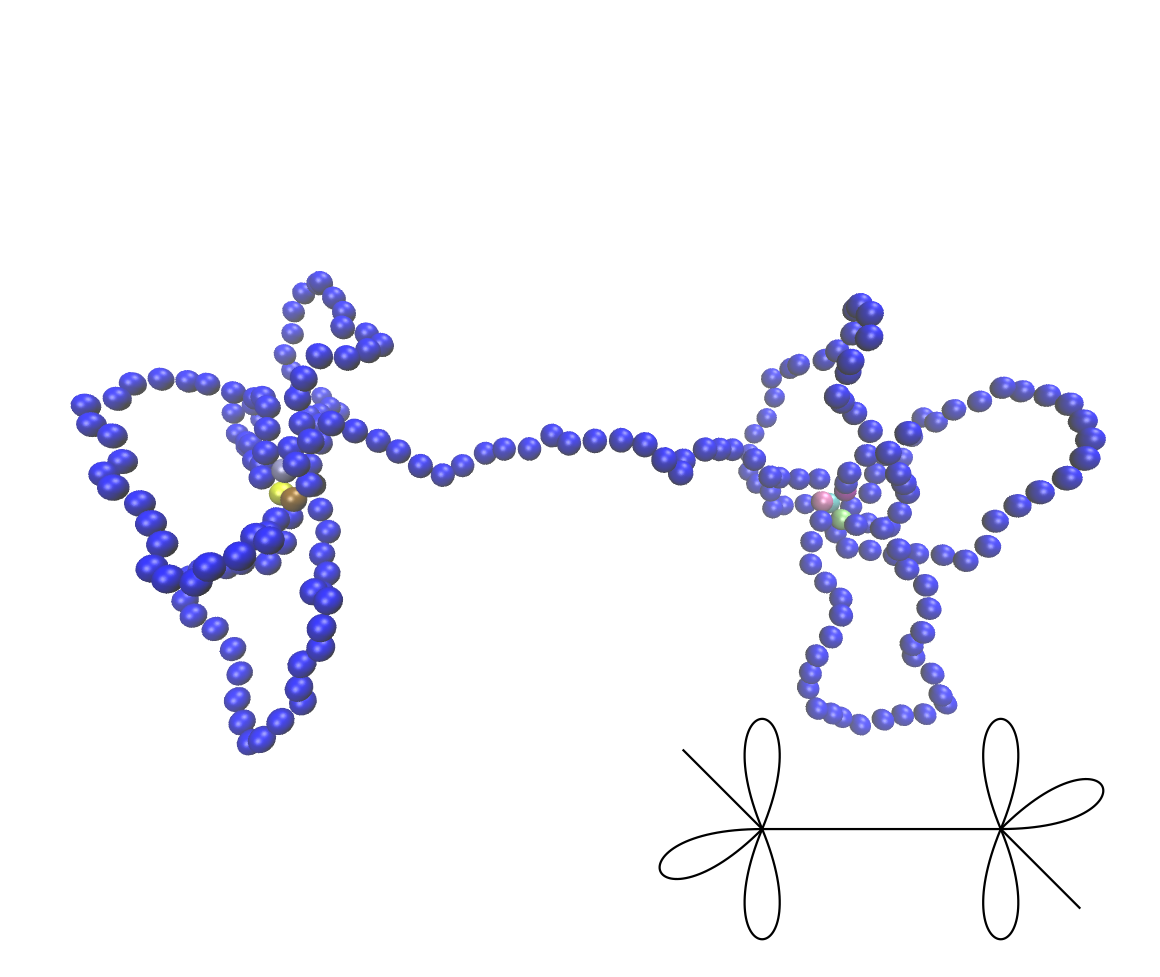}
\end{center}
\caption{Simulation snapshot and corresponding topology for the rosette case. Note different TUs are differently coloured.}
\label{rosettesnapshot}
\end{figure}

\begin{figure}
\begin{center}
\includegraphics[scale=0.25]{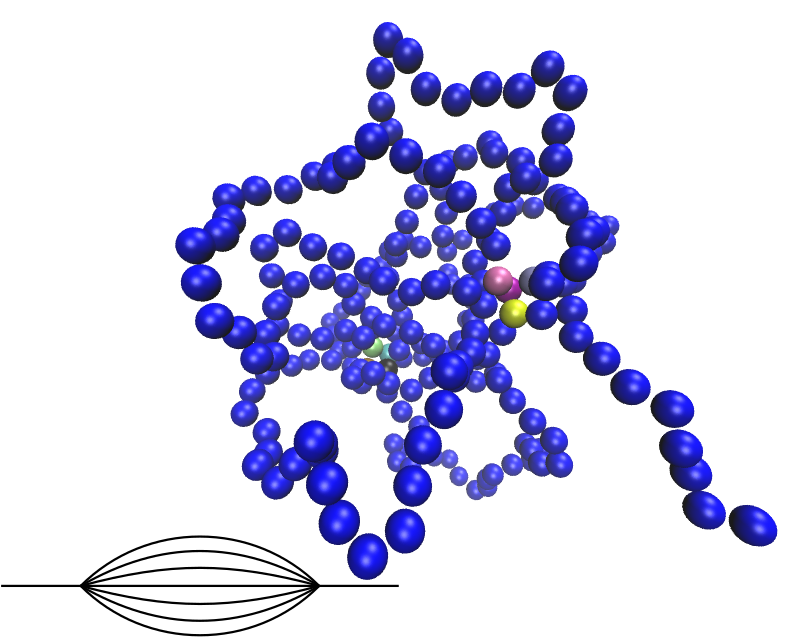}
\end{center}
\caption{Simulation snapshot and corresponding topology for the watermelon case. Note different TUs are differently coloured.}
\label{watermelonsnapshot}
\end{figure}

As expected, for small values of $\epsilon$, the chain remains unfolded. In contrast, for sufficiently large values of $\epsilon$ the targeting  topology is formed (examples of folded  configurations in this regime for the rosette and watermelon topologies are shown in Figs.~\ref{rosettesnapshot} and ~\ref{watermelonsnapshot} respectively). The point of sharpest variation of the sigmoidal curves in Fig.~\ref{r-w-transition-plot} can be interpreted as the critical interaction affinity, $\epsilon_c$, required to form the target topology (either rosette or watermelon). Thicker lines indicate the mean over $100$ random initial  configurations for each $\varepsilon$ of the normalized pairing energy ($\epairNorm$). The surrounding shaded regions represent one standard deviation on either side of this mean. 

\begin{figure}
\begin{center}
\includegraphics{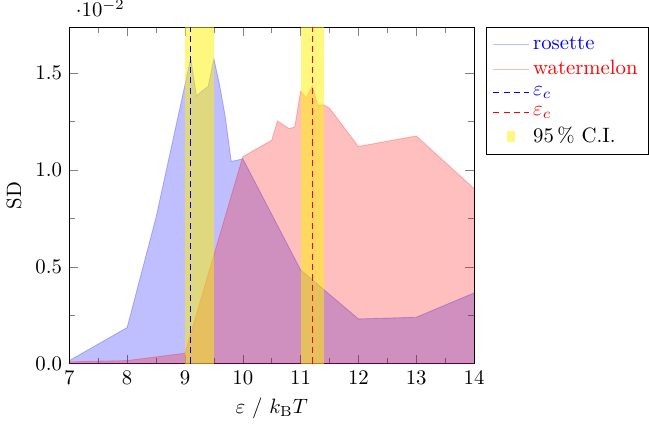}
\end{center}
\caption{Plot showing simulation results for the standard deviation of the normalised pairing energy as a function of the attraction between TU for the rosette (blue) and watermelon (red) topologies. Maxima were used to infer the transition affinities, which are indicated with dashed lines. Bootstrapped \SI{95}{\percent} confidence intervals are shaded in yellow.}
\label{sigma-plot}
\end{figure}

The interaction affinity giving rise to the maximal standard deviation is taken as the recorded transition affinity ($\epsilon_c$). Confidence intervals were formed using a bootstrapping procedure. In order to better illustrate the relationship between the standard deviation amongst simulations with the same interaction affinity, and the inferred transition affinity, the standard deviations are plotted independently in figure~\ref{sigma-plot}. From these curves, it is clear that the location of the maximum differs for the rosette and watermelon topologies. In particular, one can observe that the rosette topology forms more easily, as it requires a smaller value of $\varepsilon$, or equivalently the associated $\epsilon_c$ (corresponding to the peak in Fig.~\ref{sigma-plot}) is smaller. 

\begin{figure}
\begin{center}
\includegraphics{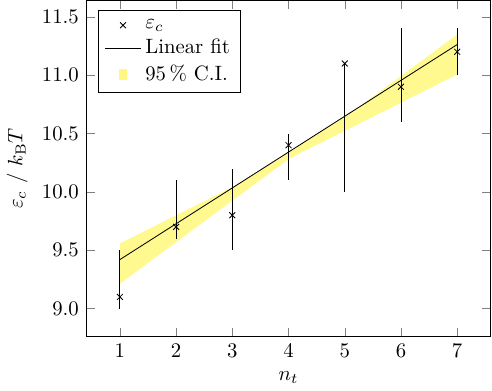}
\end{center}
\caption{Plot of critical energy, $\epsilon_c$, against number of ties $n_t$ for the first class of topologies in Table~\ref{topology-table}, with $L_{1,2}=(8,8)$. Values corresponding to $95\%$ confidence intervals for each values of $n_t$ are found by bootstrap.}
\label{affinity-fit-alpha-plot}
\end{figure}

Note that, while the rosette and watermelon topologies have the same values of $L_{1,2}$, they differ by the number of ties, $n_t$ ($n_t=0$ for the rosette topology, and $n_t=7$ for the watermelon one). In general, we observed that the larger $n_t$ in a target topology, the greater the interaction affinity typically required to form it (with exceptions, see Table~\ref{topology-table}). In this respect, a simple class to study is that of the first seven symmetric topologies shown in Table~\ref{topology-table}, of which the rosette and the watermelon constitute the limiting cases. In order to elucidate the relationship between $n_t$ and $\epsilon_c$ for this class, we carried out a bootstrapped linear regression.  The corresponding fit is plotted in Fig.~\ref{affinity-fit-alpha-plot}: a simple linear relationship between $n_t$  and $\epsilon_c$ holds to a good approximation. As the number of ties increases by one, the number of loops decreases by one too, and so our results indicate that there is a nearly uniform energetic cost each time one loop is exchanged for a tie in a chromatin network. The estimate of the constant cost per tie is $\Delta\epsilon_c = {0.31}{k_BT}$ ($95\%$ confident interval $0.24-0.36 k_BT$). 
The other two classes of topologies that are reported in Table~\ref{topology-table}) still lead to an increase of $\epsilon_c$ with $n_t$ but the functional form is less clear (see Table~\ref{topology-table} for a list of values of $\epsilon_c$ found for each inequivalent $2$-cluster topology).

\section{Topological weights of Gaussian chromatin loop networks}
Up to now, we have enumerated the configurations of polymer loop networks, thereby finding their combinatorial weights. We have also seen in the last Section that Brownian dynamic  simulations show that the energy which is required to offset the entropic cost associated with the formation of these topologies is significantly different. In the companion paper, we have additionally shown that inequivalent (unlabelled) topologies with the same combinatorial weight, such as the rosette and watermelon ones, are observed in polymer models with starkly different frequencies. In this Section, we will show that these results can be understood, at least qualitatively, by computing the {\it topological weight} of a given graph, which is essentially the partition function of a Gaussian polymer network with that topology. [Note this is equivalent to a freely-jointed polymer network with a large number of monomers~\cite{DeGennes1979}.]

More specifically, to compute the topological weight of a given graph ${\mathcal G}$, associated with an inequivalent topology of a chromatin loop network with $n$ TUs, we need to compute its corresponding partition function, 
\begin{equation}\label{generaleqweight}
Z_{\mathcal G} = \int d{\mathbf x}_1 \ldots d{\mathbf x}_n \delta({\mathcal G})
\prod_{i=0}^{n} %W_0 
e^{-\frac{3\left({\mathbf x_{i+1}}-{\mathbf x_i}\right)^2}{2l\sigma}},
\end{equation}
where $e^{-\frac{3\left({\mathbf x_{i+1}}-{\mathbf x_i}\right)^2} {2l\sigma}}$ can be thought of, in field theoretical terms, as the {\it propagator} of our Gaussian theory, from the $i$-th to the $(i+1)$-th TU. 

In the remainder of this Section, we will first compute in detail the topological weights of unlabelled configurations with $2$ clusters, which are the focus of the numerical simulations in the companion paper~\cite{short}. Afterward, we shall see how to generalise the calculation to compute the topological weight of any given Gaussian polymer loop network. This calculation can be done explicitly because we are approximating the polymer with a Gaussian chain. Including self-avoidance and mutual avoidance between different polymer, segments would require a separate treatment and is outside the scope of the current work. In the special case of $2$-cluster configurations, self-avoidance is included in the discussion of the results in~\cite{short}.

\subsection{Topological weights of $2$-cluster configurations}

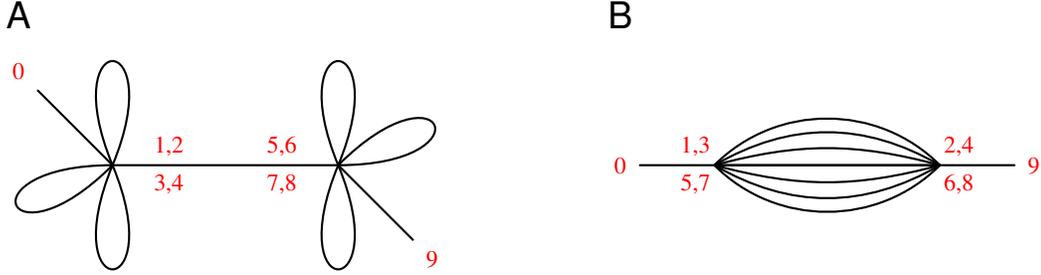
\begin{figure}

\centering{

\begin{tikzpicture}[scale=1.00]
    \SetGraphUnit{3}
    \Vertex{A}
    \EA(A){B}
    \NOWE[unit=1](A){in}
    \SOEA[unit=1](B){out}
    \Edge(A)(B)
    \Edge(in)(A)
    \Edge(B)(out)
    \loopNarrowN{A}
    \loopNarrowS{A}
    \loopNarrowWSW{A}
    \loopNarrowN{B}
    \loopNarrowS{B}
    \loopNarrowENE{B}

    \EA[unit=4](B){in2}
    \EA[unit=1](in2){C}
    \EA(C){D}
    \EA[unit=1](D){out2}
    \Edge(C)(D)
    \Edge(in2)(C)
    \Edge(C)(out2)
    \foreach \angle in {15,30,45}
        {
        \tikzset{EdgeStyle/.append style = {bend left = \angle}}
        \Edge(C)(D)
        \Edge(D)(C)
        };

    \node (note) at (-1.25,2) {{\fontsize{14}{0pt}\usefont{T1}{phv}{m}{n} A}};
    \node (note) at (6.75,2) {{\fontsize{14}{0pt}\usefont{T1}{phv}{m}{n} B}};

    \node (note) at (-1.25,1.25) {{\textcolor{red}{0}}};
    \node (note) at (0.75,0.25) {{\textcolor{red}{1,2}}};
    \node (note) at (0.75,-0.25) {{\textcolor{red}{3,4}}};
    \node (note) at (2.25,0.25) {{\textcolor{red}{5,6}}};
    \node (note) at (2.25,-0.25) {{\textcolor{red}{7,8}}};
    \node (note) at (4.25,-1.25) {{\textcolor{red}{9}}};

    \node (note) at (6.75,0) {{\textcolor{red}{0}}};
    \node (note) at (7.75,0.25) {{\textcolor{red}{1,3}}};
    \node (note) at (7.75,-0.25) {{\textcolor{red}{5,7}}};
    \node (note) at (11.25,0.25) {{\textcolor{red}{2,4}}};
    \node (note) at (11.25,-0.25) {{\textcolor{red}{6,8}}};
    \node (note) at (12.25,0) {{\textcolor{red}{9}}};
    
\end{tikzpicture}
}
\caption{Loop network configurations and TU labelling used for the calculation of the topological weights of the rosette (A) and watermelon (B) topologies.}\label{topologicalweightRW}
\end{figure}

We begin by noting that the term $\delta({\mathcal G})$ in Eq.~(\ref{generaleqweight}) is a product of Dirac $\delta$ functions which specify the topology of the network~\cite{Duplantier1989}. For instance, in the case of rosettes (${\mathcal G\equiv \mathcal R}$, Fig.~\ref{topologicalweightRW}A) and watermelons (${\mathcal G\equiv \mathcal W}$, Fig.~\ref{topologicalweightRW}B), $\delta({\mathcal G})$ is explicitly given by $\delta({\mathcal R})$ and $\delta({\mathcal W})$, with
\begin{eqnarray}
\delta({\mathcal R}) & = & \prod_{i=2,3,4}\delta({\mathbf{x_1}}-{\mathbf{x_i}}) 
\prod_{j=6,7,8} \delta({\mathbf{x_5}}-{\mathbf{x_j}}) \\ \nonumber
\delta({\mathcal W}) & = & \prod_{i=3,5,7}\delta({\mathbf{x_1}}-{\mathbf{x_i}}) \prod_{j=4,6,8}\delta({\mathbf{x_2}}-{\mathbf{x_j}}).
\end{eqnarray}

The topological weight of the rosette is therefore given by
\begin{equation}\label{rosetteweight}
Z_{\mathcal R} = \int d{\mathbf x}_0 \ldots d{\mathbf x}_9  
\left[\prod_{i=0}^{8} %W_0 
e^{-\frac{3\left({\mathbf x_{i+1}}-{\mathbf x_i}\right)^2}{2l\sigma}}\right]
\prod_{i=2,3,4}\delta({\mathbf{x_1}}-{\mathbf{x_i}}) 
\prod_{j=6,7,8} \delta({\mathbf{x_5}}-{\mathbf{x_j}}),
\end{equation}
where the TU labelling in the integral follows the one in Fig.~\ref{topologicalweightRW}A.

Noting that
\begin{equation}
 \int d{\mathbf x}_0 %W_0 
 e^{-\frac{3\left({\mathbf x_{1}}-{\mathbf x_0}\right)^2}{2l\sigma}} = \int d{\mathbf x}_0 %W_0 
 e^{-\frac{3{\mathbf x_0}^2}{2l\sigma}}= %W_0
 W_0^{-1},%=1,
\end{equation}
with $W_0=\left(\frac{3}{2\pi l\sigma}\right)^{3/2}$,
and that an analogous formula holds for the integral over $d{\mathbf x_9}$, we obtain, by making use of the properties of the $\delta$ function, that
\begin{eqnarray}\label{rosetteweight2}
Z_{\mathcal R} & = & \int d{\mathbf x}_1 d{\mathbf x}_5  
W_0^{-2} e^{-\frac{3\left({\mathbf x_{1}}-{\mathbf x_5}\right)^2}{2l\sigma}} \\ \nonumber
& = & W_0^{-3} V,
\end{eqnarray}
where we have called $V$ the volume of the system.

By repeating the same steps for the watermelon topology (see Fig.~\ref{topologicalweightRW}B, and the associated choice of TU labelling), we get
\begin{eqnarray}\label{rosetteweight3}
Z_{\mathcal W} & = & \int d{\mathbf x}_1 d{\mathbf x}_5  
W_0^{-2} e^{-7\frac{3\left({\mathbf x_{1}}-{\mathbf x_5}\right)^2}{2l\sigma}} \\ \nonumber
& = & \frac{W_0^{-3} V}{7^{3/2}} = \frac{Z_{\mathcal R}}{7^{3/2}}.
\end{eqnarray}

Therefore, the topological weight of the watermelon is much smaller than that of the rosette. Additionally, one can generalise the result shown above to hybrid rosette-watermelon configurations with $2$ clusters and $n_t$ ties, obtaining that their topological weight is given by
\begin{equation}
Z = \frac{Z_{\mathcal R}}{n_t^{3/2}},
\end{equation}
which becomes Eq.~(\ref{rosetteweight3}) for $n_t=7$ (which holds for the watermelon topology). The decrease in topological weight of $2$-cluster topologies with $n_t$ qualitatively explains why they are seen less frequently in simulations~\cite{short}, and why the interaction energy between TU needed to stabilise a topology increases with $n_t$, as found in the previous Section with coarse-grained molecular dynamics simulations.

\subsection{General formulas for the topological weights of Gaussian loop networks}

With a bit more work, the topological weight calculation just outlined can actually be generalised to any chromatin loop network.

\if{\begin{tikzpicture}[scale=1.00]
    \SetGraphUnit{3}
    \Vertex{A}
    \EA(A){B}
    \WE[unit=1](A){in}
    \EA[unit=1](B){out}
    \Edge(A)(B)
    \Edge(in)(A)
    \Edge(B)(out)
    \foreach \angle in {15,30,45}
        {
        \tikzset{EdgeStyle/.append style = {bend left = \angle}}
        \Edge(A)(B)
        \Edge(B)(A)
        };
\end{tikzpicture}}\fi

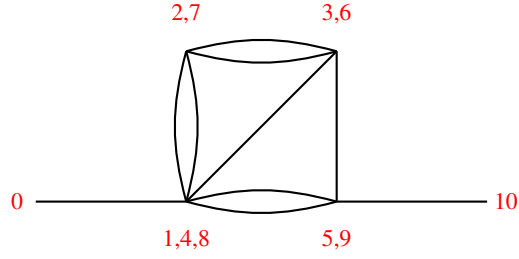
\begin{figure}
\begin{tikzpicture}[scale=1.00]
    \SetGraphUnit{3}
    \Vertex{A}
    \NO[unit=2](A){B}
    \WE[unit=2](A){in}
    \EA[unit=2](B){C}
    \SO[unit=2](C){D}
    \EA[unit=2](D){out}
    {
        \tikzset{EdgeStyle/.append style = {bend left = 15}}
        \Edge(A)(B)
        \Edge(B)(C)
        \Edge(A)(D) 
        \tikzset{EdgeStyle/.append style = {bend left = -15}}
        \Edge(A)(B)
        \Edge(B)(C)
        \Edge(A)(D)
    };
    \Edge(in)(A)
    \Edge(C)(A)
    \Edge(C)(D)
    \Edge(D)(out)
    \node (note) at (-2.25,0) {{\textcolor{red}{0}}};
    \node (note) at (0,-0.5) {{\textcolor{red}{1,4,8}}};
    \node (note) at (0,2.5) {{\textcolor{red}{2,7}}};
    \node (note) at (2,2.5) {{\textcolor{red}{3,6}}};
    \node (note) at (2,-0.5) {{\textcolor{red}{5,9}}};
    \node (note) at (4.25,0) {{\textcolor{red}{10}}};
\end{tikzpicture}
\caption{Loop network configuration and TU labelling used for the calculation of the topological weights of a network with multiple clusters (here $4$).}\label{topologicalweightgenerictopology}
\end{figure}

To see how let us consider the topology ${\mathbf G}$ shown in Fig.~\ref{topologicalweightgenerictopology}. Its associated topological weight is given by:
\begin{equation}\label{generictopologyweight}
Z_{\mathcal G} = \int d{\mathbf x}_0 \ldots d{\mathbf x}_{10}  
\left[\prod_{i=0}^{9} %W_0 
e^{-\frac{3\left({\mathbf x_{i+1}}-{\mathbf x_i}\right)^2}{2l\sigma}}\right]
\delta({\mathbf{x_1}}-{\mathbf{x_4}})
\delta({\mathbf{x_1}}-{\mathbf{x_8}})
\delta({\mathbf{x_2}}-{\mathbf{x_7}})
\delta({\mathbf{x_3}}-{\mathbf{x_6}})
\delta({\mathbf{x_5}}-{\mathbf{x_9}}),
\end{equation}
which, by using methods similar to those described in the previous section,  can also be written as,
\begin{eqnarray}\label{generictopologyweight2}
Z_{\mathcal G} & = & \int d{\mathbf x}_1 d{\mathbf x}_{2} d{\mathbf x}_3 d{\mathbf x}_5\,  W_0^{-2}\, e^{-\frac{3}{2l\sigma} f({\mathbf x}_1, {\mathbf x}_2, {\mathbf x}_3, {\mathbf x}_5)} \\ \nonumber
f({\mathbf x}_1, {\mathbf x}_2, {\mathbf x}_3, {\mathbf x}_5) & = & 
\left[2\left({\mathbf x_{2}}-{\mathbf x_1}\right)^2+2\left({\mathbf x_{3}}-{\mathbf x_2}\right)^2+\left({\mathbf x_{3}}-{\mathbf x_1}\right)^2+2\left({\mathbf x_{5}}-{\mathbf x_3}\right)^2+2\left({\mathbf x_{5}}-{\mathbf x_1}\right)^2\right] .
\end{eqnarray}

We now introduce the following matrix, 
\begin{equation}
	A(\mathcal{G})=
	\begin{pmatrix}
		0 & 2 & 1 & 2 \\
		2 & 0 & 2 & 0 \\
		1 & 2 & 0 & 1 \\
        2 & 0 & 1 & 0
	\end{pmatrix},
\end{equation}
equal to the adjacency matrix of the multi-graph corresponding to ${\mathbf G}$ (note that, if loops were present in ${\mathbf G}$, they should not be included in the calculation of $A(\mathcal{G})$). From this, we define the  matrix,
\begin{equation}
	B(\mathcal{G})=
	\begin{pmatrix}
		5 & -2 & -1 & -2 \\
		-2 & 4 & -2 & 0 \\
		-1 & -2 & 4 & -1 \\
        -2 & 0 & -1 & 3
	\end{pmatrix},
\end{equation}
where we have changed the sign of the off-diagonal components and  the diagonal components have been set equal to the sum of the corresponding row in $A(\mathcal{G})$. With this setup the argument of the exponential in Eq.~(\ref{topologicalweightgenerictopology}) can be written in matrix form as, 
\begin{equation}
f({\mathbf x})={\mathbf{x}}^T B({\mathcal G}) {\mathbf{x}},
\end{equation}
where $\mathbf{x}^T=({\mathbf x}_1, {\mathbf x}_2,{\mathbf x}_3,{\mathbf x}_5)$. Note that $\det(B(\mathbf G))=0$.  This is consistent with the fact that the topological weight is proportional to the volume $V$ of the system. By fixing  the position of one of the cluster centre of mass, say ${\mathbf x}_1$, and integrating over it, the  weight associated with this topology can be given in terms of the determinant of the matrix obtained by removing the first row and column, 
\begin{equation}
	\det(B'(\mathcal{G}))=\det
	\begin{pmatrix}
		4 & -2 & 0 \\
		-2 & 4 & -1 \\
        0 & -1 & 3
	\end{pmatrix} =32,
\end{equation}
as follows
\begin{equation}
Z_{\mathcal G}=W_0^{-3}\, V \det(B'(\mathcal{G}))^{-3/2}=\frac{W_0^{-3}\, V}{32^{3/2}},
\end{equation}

It can be verified that, as expected, the above result  does not depend on  which cluster is fixed and integrated upon, as the determinant of any matrix obtained by removing the $i$-th row and column is the same. 

By applying this  procedure, for instance,  to a string of rosettes with  $n$ TUs one can show that the corresponding topological weight is 
$W_0^{-3}V$. 

Finally for  a  general graph ${\mathbf G}$ with $n$ TUs and $k$ clusters, its topological weight can be computed by starting from the corresponding matrix $B({\mathcal G})$ and computing the determinant of any sub-matrix $B'({\mathcal G})$ obtained by removing the $i$-th row and column, for any $i\in [1,k]$. This is given by
\begin{equation}
Z_{\mathcal G}=W_0^{-3} V \det(B'(\mathcal{G}))^{-3/2}.
\end{equation}
The above result confirms that a generic network typically has a (significantly) lower weight with respect to that of a string of rosettes with the same number of TUs $n$. This is in line with the numerical results obtained for $k=2$.

\section{Discussion and conclusions}

In summary, we have presented a combination of analytical and numerical results for the combinatorial and topological weights of chromatin loop networks. These weights are important to determine the relative frequencies with which different topologies arise in polymer models for DNA and chromatin, which are studied in the companion paper~\cite{short}. In particular, we are interested here in the relation between these results and the physical properties of the loop networks which arise due to the bridging-induced attraction~\cite{Brackley2013,Brackley2016,Brackley2021}, in polymer models for the 3D structures formed by chromatin fibres {\it in vivo}, and which are associated with gene folding. For instance, the statistical, or Boltzmann, weight associated with a given topology shown in Table~\ref{topology-table}, which determines the frequency with which it is observed for instance in computer simulations, is proportional to its combinatorial weight (computed in Section III) times its topological weight (computed in Section V). 

We have shown that the enumeration problems associated with counting labelled and unlabelled chromatin loop networks are fundamentally different. When transcription units (TUs) are labelled (Section II), the problem can be usefully mapped to that of counting the ways in which $n$ different TUs can be distributed into $k$ clusters with $\ge 2$ TUs per cluster. The resulting combinatorial sequences are often related to the Bell or Stirling numbers, and we have shown that it is possible to find explicit formulas for the exponential generating functions associated with a number of different cases, with or without singletons (i.e., TUs not in any clusters). This is useful for providing estimates or upper bounds for the number of topologies which a given chromatin region (with a specified number of TUs) can fold into. 

For networks with unlabelled TUs, corresponding to inequivalent topologies (Section III), the enumeration problem is related to that of counting multi-graph, which is $NP$-complete, hence harder. We have though provided here a derivation of a formula counting all inequivalent topologies with $n$ TUs and $k=2$ clusters; for $n=8$ (a common occurrence in real gene loci~\cite{Chiang2022b}) and $k=2$, the case studied in detail in the continuum paper, this formula gives $20$ inequivalent topologies (shown explicitly in Table~\ref{topology-table}). 

We also asked what attraction energy is needed to form a target topology. This is a biophysically relevant question as regards chromatin loop networks: for instance, we may want to know whether a rosette topology or a watermelon one forms more easily (i.e., requires less interaction between the TUs), as this may affect the relative frequency with which these two structures are observed either in computer simulations of chromatin fibres~\cite{short,Chiang2022b}, or experimentally in 3D structures of gene loci~\cite{Dotson2022}. Previous work based on renormalisation group calculations came to the important conclusion that the entropic exponent of a polymer loop network solely depends on the degree of its nodes (the clusters in our terminology)~\cite{Duplantier1986,Duplantier1989}. However, this exponent does not completely determine the weight, as there is a pre-factor which could in principle be also topology-dependent. More in detail, rosettes and watermelons, and indeed all first seven network topologies in Table~\ref{topology-table}, have the same entropic exponent, yet they require significantly different energies to form, as we show in Section IV. In particular, by focussing on configurations with $n=8$ TUs and $k=2$ clusters, our simulations show that the critical energy to form a target topology tends to increase with the number of ties, or polymer segments, linking the two clusters, which we call $n_t$. 

In Section V, we compute the topological weight of a chromatin loop network, under the assumption that the polymer is a Gaussian chain. This weight is the partition function of a network with the given topology and, importantly, we find that it depends strongly on $n_t$, qualitatively explaining our numerical results in Section IV. 

In the future, it would be interesting to generalise the topological weight calculations in Section V to the case where the polymer network has both self- and mutual avoidance. From an application perspective, it would be desirable to use our labelled and inequivalent unlabelled topologies to classify the 3D configurations of chromatin fibre around genes, for instance, the gene loci configuration found by ``HiP-HoP'' simulations in~\cite{Chiang2022b}, or the interaction networks and hypergraphs found by chromatin capture experiments accounting for multiway chromatin contacts, such as poreC~\cite{Dotson2022}. We hope that these extensions of our work will be addressed in the future.

 This work was supported by the Wellcome Trust (223097/Z/21/Z). For the purpose of open access, the authors have applied a Creative Commons Attribu- tion (CC BY) licence to any Author Accepted Manuscript version arising from this submission. 

%\bibliographystyle{apsrev4-1}
%\bibliography{references}


\end{document}